\documentclass[preprint,aps,prc,showpacs,footinbib,secnumarabic]{revtex4-1}

\usepackage{graphicx}% Include figure files
\usepackage{bm}% bold math
\usepackage{color}
\usepackage{hyperref}
\newcommand{\bvec}[1]{\ensuremath{\boldsymbol{#1}}}

\begin{document}

\title{Polarized $\chi_{c2}$-charmonium production
       in antiproton-nucleus interactions}

\author{A.B. Larionov$^{1,2}$\footnote{Corresponding author.\\ 
        E-mail address: larionov@fias.uni-frankfurt.de},
        M. Strikman$^3$, M. Bleicher$^{1,4}$}

\affiliation{$^1$Frankfurt Institute for Advanced Studies (FIAS), 
             D-60438 Frankfurt am Main, Germany\\ 
             $^2$National Research Center "Kurchatov Institute", 
             123182 Moscow, Russia\\
             $^3$Pennsylvania State University, University Park, PA 16802, USA\\
             $^4$Institut f\"ur Theoretische Physik, J.W. Goethe-Universit\"at,
             D-60438 Frankfurt am Main, Germany}

\begin{abstract}
Starting from the Feynman diagram representation of multiple scattering we consider
the polarized $\chi_c$(1P)-charmonia production in antiproton-nucleus reactions
close to the threshold ($p_{\rm lab}=5-7$ GeV/c). The rescattering and absorption of
the incoming antiproton and outgoing charmonium on nucleons are taken into account,
including the possibility of the elastic and  nondiagonal (flavor-conserving)
scattering $\chi_{cJ} N \to \chi_{cJ^\prime} N$, $J,J^\prime=0,1,2$. 
The elementary amplitudes of the latter 
processes are evaluated by expanding the physical $\chi_c$-states in the Clebsch-Gordan
series of the $c \bar c$ states with fixed values of internal orbital angular momentum
($L_z$) and spin projections on the $\chi_c$ momentum axis. The total interaction cross
sections of these $c \bar c$
states with nucleons have been calculated in previous works using the QCD factorization
theorem and the nonrelativistic quarkonium model and turned out to be strongly $L_z$-dependent
due to the transverse size difference. This directly leads to finite values of the
$\chi_c$-nucleon nondiagonal scattering amplitudes. We show that the $\chi_{c0} N \to \chi_{c2} N$
transitions significantly influence the $\chi_{c2}$-production with helicity 
zero at small transverse momenta. This can serve as a signal in future experimental
tests of the quark structure of $\chi_c$-states by the PANDA collaboration at FAIR. 
\end{abstract}

\pacs{25.43.+t;~14.40.Pq;~24.10.Ht}

\maketitle

\section{Introduction}

It is well established, that in the perturbative QCD regime, $r_t \to 0$, the 
total cross section of a quarkonium state interaction with a proton scales 
as the square of the transverse separation $r_t$ between quark and antiquark,
$\sigma_{\bar q q}(r_t) \propto r_t^2$. This indicates that extracting the 
total quarkonium-nucleon cross section gives access to the transverse size
of the quarkonium, although at $r_t > 0.2-0.5$ fm the deviations from 
a simple proportionality become important (the energy dependence of the 
dipole-nucleon cross section also modifies this relation). If the relative 
coordinate wave function of the quarkonium is nonisotropic (P,D,...states),
it is, thus, natural to expect that the cross section will depend on the 
quarkonium polarization.

This was first predicted in Ref. \cite{Gerland:1998bz}, where the cross sections 
of the charmonium- and bottomonium-nucleon interaction have been calculated
on the basis of the QCD factorization theorem and the nonrelativistic quarkonium 
model. Indeed, for the 1P $\chi_c$- and $\chi_b$-states this resulted in a quarkonium 
polarization dependent total interaction cross section with a nucleon. 
Qualitatively similar results were obtained later in Ref. \cite{Hufner:2000jb}, however,
with somewhat different absolute values of the charmonium-nucleon cross sections.  

Analyzing charmonium production in relativistic heavy-ion collisions, 
the authors of Ref. \cite{Gerland:1998bz} predicted,
that the survival probabilities of $\chi$ states with different polarization
will, therefore, be different. This color filtering effect has been included
afterwards in dynamical UrQMD simulations of heavy-ion collisions
\cite{Spieles:1999pm} to successfully describe $J/\psi$ production at SPS energies.
Unfortunately, heavy-ion collisions involve too complex processes and it is difficult 
to use them to access the true charmonium-nucleon cross sections \cite{Vogt:1999cu}.

Antiproton-nucleus collisions give the unique opportunity to study nuclear
interactions of the slowly moving charmonium states exclusively formed in 
$\bar p p \to \Psi$ reactions inside the nuclear medium \cite{Brodsky:1988xz,Farrar:1989vr}.
Here, $\Psi$ stands for any charmonium state ($J/\psi,~\psi^\prime,~\chi_c$...)
decaying to $\bar p p$.  
In this paper we show that owing to the polarization-dependent $\chi_c$-nucleon
cross sections the produced $\chi_{c2}$ states in near-threshold $\bar p$-nucleus
collisions should reveal a significant polarization signal. Complementary information
can be obtained in $\gamma A \to J/\psi A^*$ reactions at $E_{\gamma} \sim 10$ GeV 
which will be studied at the upgraded TJNAF facility.

We calculate the Feynman multiple scattering diagrams in the generalized eikonal 
approximation (GEA) \cite{Frankfurt:1996xx,Sargsian01}. The direct formation
mechanism $\bar p p \to \chi_c$, as well the corrections due to the
rescattering of incoming antiproton and outgoing charmonium states on nucleons,
including the possibility of nondiagonal transitions, are taken into account.
The nondiagonal transitions $\chi_{J_1\nu}N \to \chi_{J\nu}N$ 
are easily possible due to the small ($\sim 140$ MeV) mass splitting between
the various $\chi_c$ states.
We show that the nondiagonal transitions strongly enhance the polarization signal
with respect to the color filtering mechanism only. 

In sec. \ref{model} we describe our model. 
Section \ref{results} contains the results of the numerical 
calculations for the transverse momentum differential cross sections
of $\chi_c$ production with different total angular momenta and
helicities. At the end of sec. \ref{results} we propose concrete
signals for the future PANDA experiment at FAIR. Section \ref{summary}
summarizes the main results of this work. 
Appendix \ref{MultScatt}
contains the derivation of the expressions for the multiple scattering
amplitudes.    

\section{Model}
\label{model}

In the following for brevity we denote as $\chi_J$ the $\chi_{cJ}$
charmonium with the total angular momentum $J$ ($J=0,1,2$).
When explicitly needed, we will also use the notation 
$\chi_{J\nu}$ for the $\chi_{cJ}$ states with the fixed helicity $\nu$ 
($\nu=-J, \ldots ,J$).

\begin{figure}
\includegraphics[scale = 0.7]{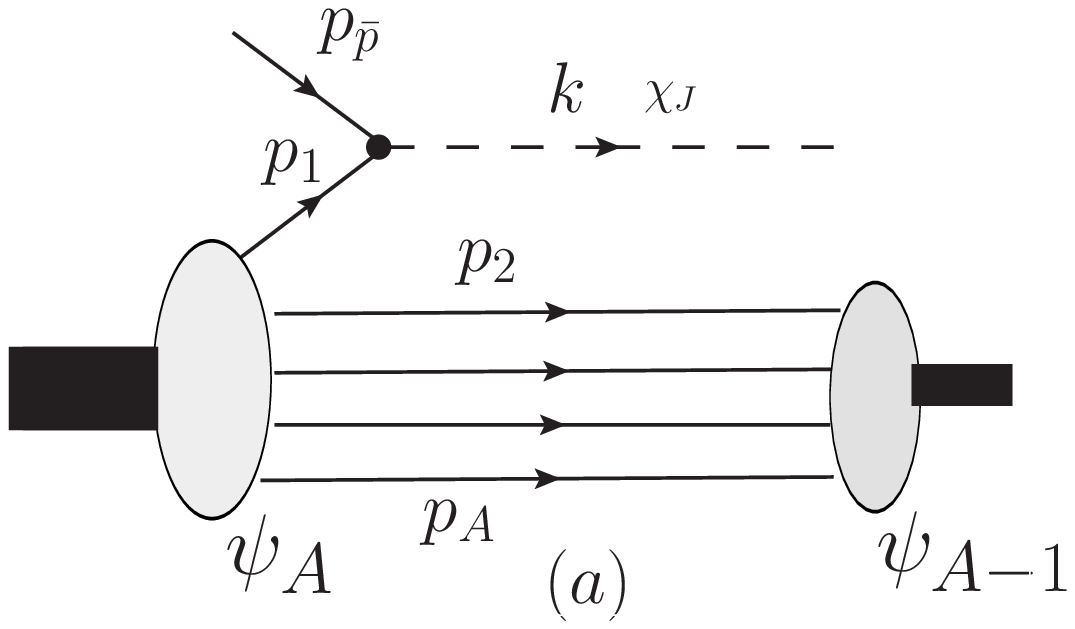}
\includegraphics[scale = 0.7]{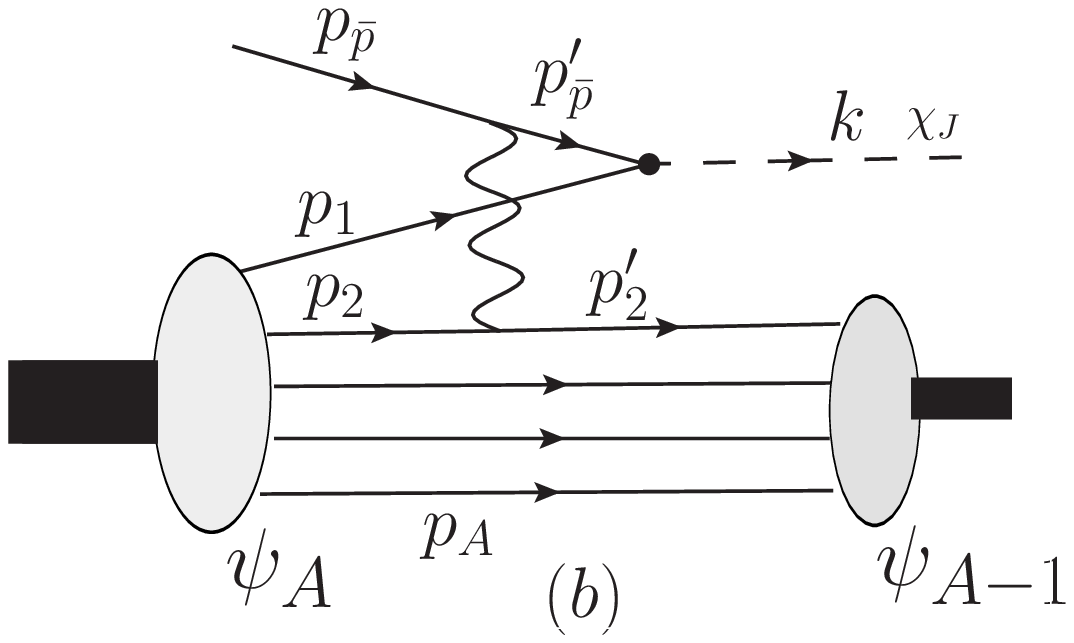}
\includegraphics[scale = 0.7]{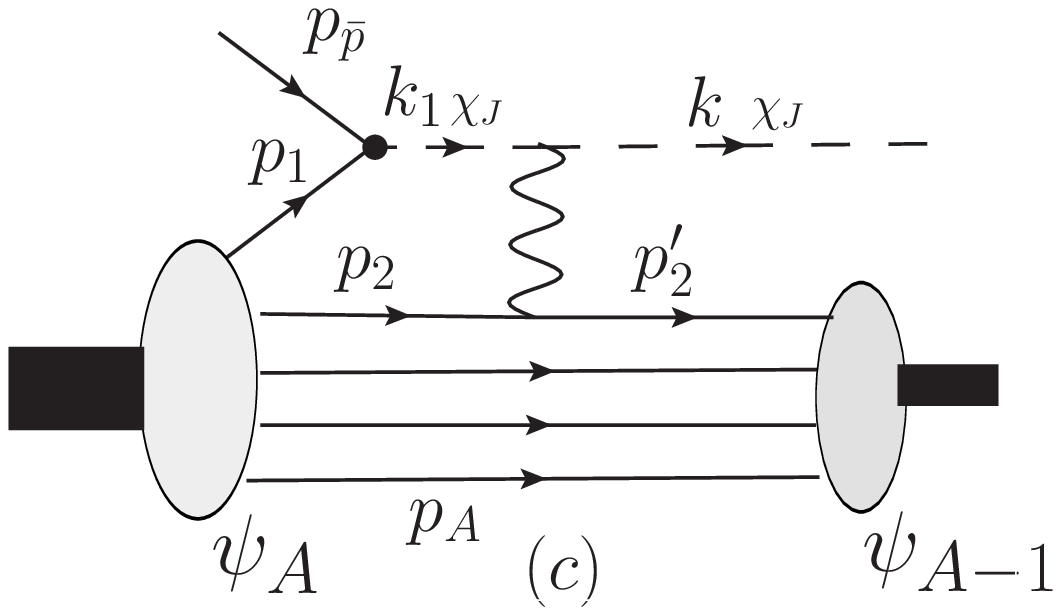}
\includegraphics[scale = 0.7]{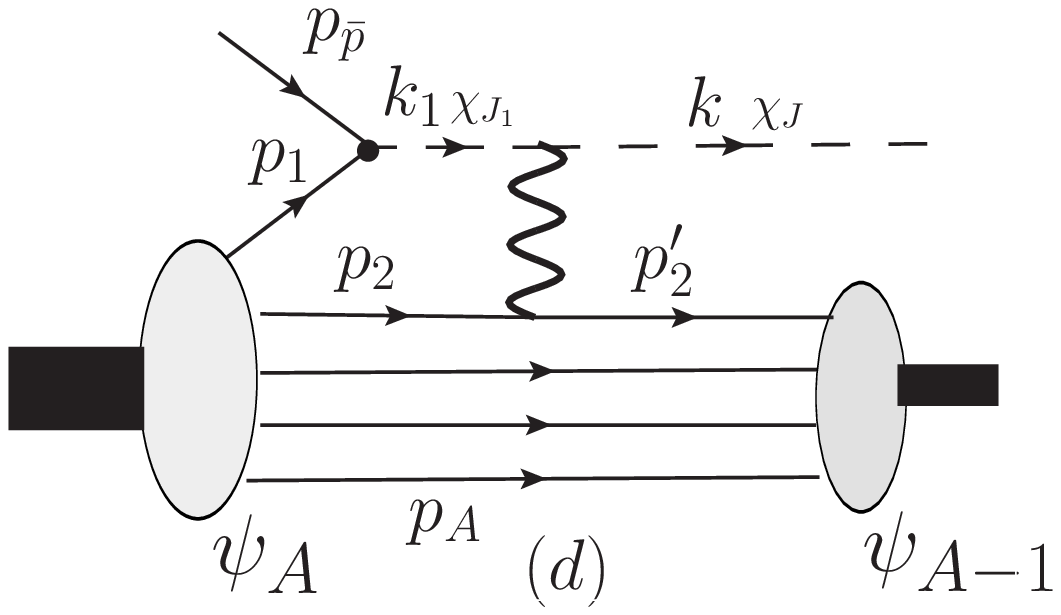}
\caption{\label{fig:pbarChi_dia} (a) The diagram for the production of 
the charmonium state $\chi_J$ with four-momentum $k$ in the impulse
approximation. (b) The diagram taking into account elastic rescattering
of the incoming antiproton on a nucleon. (c) The diagram with rescattering  
$\chi_J N_2 \to \chi_J N_2^\prime$ of the initially 
produced $\chi_J$-state on a nucleon. 
(c) The diagram with the initial production of another 
state  $\chi_{J_1}$ followed by the nondiagonal transition
$\chi_{J_1} N_2 \to \chi_{J} N_2^\prime$. $p_{\bar p}^\prime$,
$k_1$ and $p_2^\prime$ are the four-momenta of the intermediate 
antiproton and charmonium states and of the scattered
nucleon $N_2^\prime$, respectively.}
\end{figure}

Let us first consider only one- and two-step reactions.
In this approximation, all possible diagrams contributing 
to the exclusive process ${\bar p} A \to \chi_J (A-1)^*$ are shown in 
Fig.~\ref{fig:pbarChi_dia}.
We neglect the contribution of the processes where $\bar p$ 
first excites to $\bar N^*$ and next the reaction  
$\bar N^* + p \to \chi_J$ takes place. This should be a reasonable assumption
since at the beam momentum of 5.7 GeV/c the diffractive cross section 
$\sigma(\bar p p \to \bar N^* p + \mbox{c.c.})=0.13\pm0.02$ mb 
\cite{Atherton:1976jy} is two orders of magnitude smaller than 
the elastic $\bar p p$ cross section ($\simeq 15$ mb) at the same 
beam momentum. (Another reason is that the Dalitz plots for the 
$\chi_c \to \bar p p \pi^0$ decay reported by CLEO \cite{Onyisi:2010nr} 
do not show any structures at $M_{\bar p \pi^0}^2 \simeq 2$ GeV$^2$ or 
at $M_{p \pi^0}^2 \simeq 2$ GeV$^2$. Hence the $\chi_c$ coupling to 
the $\bar N^* N$ (+c.c.) states is not expected to be significant.)
The amplitudes for the processes (a),(b),(c) and (d) are, respectively
\begin{eqnarray}
      M^J(1) &=& 
      \frac{M_{J; {\bar p}p}(\mathbf{k}-\mathbf{p}_{\bar p})}{\sqrt{2E_1}}
   \int d^3x_1 ...d^3x_A\psi^*_{A-1}(\mathbf{x}_2,...,\mathbf{x}_A) \nonumber \\
   &\times& \mbox{e}^{-i(\mathbf{k}-\mathbf{p}_{\bar p})\mathbf{x}_1}
   \psi_{A}(\mathbf{x}_1,\mathbf{x}_2,...,\mathbf{x}_A)~,     \label{M^J(1)} \\
      M^J(2,1) &=&
      \frac{1}{\sqrt{2E_1}2E_2(2\pi)^6} \int d^3x_2^\prime d^3x_1 ...d^3x_A
      \psi^*_{A-1}(\mathbf{x}_2^\prime,\mathbf{x}_3,...,\mathbf{x}_A) \nonumber \\
   &\times& \psi_{A}(\mathbf{x}_1,\mathbf{x}_2,...,\mathbf{x}_A)
    \int d^3p_2^\prime d^3p_1 d^3p_2 
         \mbox{e}^{i\mathbf{p}_2^\prime\mathbf{x}_2^\prime 
                 - i\mathbf{p}_1\mathbf{x}_1 - i\mathbf{p}_2\mathbf{x}_2} \nonumber \\
   &\times& \delta^{(3)}(\mathbf{p}_{\bar p}+\mathbf{q}_2+\mathbf{p}_1
                         -\mathbf{k})
            \frac{ M_{J; {\bar p}p}(\mathbf{k}-\mathbf{p}_{\bar p}^\prime)
                   M_{\bar p N}(\mathbf{q}_2) 
                 }{D_{\bar p}(p_{\bar p}^\prime)}~,      \label{M^J(2,1)}\\
      M^J(1,2) &=&
      \frac{1}{\sqrt{2E_1}2E_2(2\pi)^6} \int d^3x_2^\prime d^3x_1 ...d^3x_A
      \psi^*_{A-1}(\mathbf{x}_2^\prime,\mathbf{x}_3,...,\mathbf{x}_A) \nonumber \\
   &\times& \psi_{A}(\mathbf{x}_1,\mathbf{x}_2,...,\mathbf{x}_A)
    \int d^3p_2^\prime d^3p_1 d^3p_2 
         \mbox{e}^{i\mathbf{p}_2^\prime\mathbf{x}_2^\prime 
                 - i\mathbf{p}_1\mathbf{x}_1 - i\mathbf{p}_2\mathbf{x}_2} \nonumber \\
   &\times& \delta^{(3)}(\mathbf{p}_{\bar p}+\mathbf{p}_1
                        +\mathbf{q}_2-\mathbf{k})
         \frac{M_{JN}(\mathbf{q}_2) 
         M_{J; {\bar p}p}(\mathbf{k}_1-\mathbf{p}_{\bar p})}{D_J(k_1)}~,\label{M^J(1,2)}\\
      M^{J_1J}(1,2) &=&
      \frac{1}{\sqrt{2E_1}2E_2(2\pi)^6} \int d^3x_2^\prime d^3x_1 ...d^3x_A
      \psi^*_{A-1}(\mathbf{x}_2^\prime,\mathbf{x}_3,...,\mathbf{x}_A) \nonumber \\
   &\times& \psi_{A}(\mathbf{x}_1,\mathbf{x}_2,...,\mathbf{x}_A)
    \int d^3p_2^\prime d^3p_1 d^3p_2 
         \mbox{e}^{i\mathbf{p}_2^\prime\mathbf{x}_2^\prime 
                 - i\mathbf{p}_1\mathbf{x}_1 - i\mathbf{p}_2\mathbf{x}_2} \nonumber \\
   &\times& \delta^{(3)}(\mathbf{p}_{\bar p}+\mathbf{p}_1
                        +\mathbf{q}_2-\mathbf{k})
           \frac{M_{JN^\prime;J_1N}(\mathbf{q}_2) 
   M_{J_1; {\bar p}p}(\mathbf{k}_1-\mathbf{p}_{\bar p})}{D_{J_1}(k_1)}~,  \label{M^J1J(1,2)}  
\end{eqnarray}
where $q_2=p_2-p_2^\prime$ is the four-momentum transfer from the nucleon-scatterer.
Here $E_1$ and $E_2$ are the single-particle energies of the involved nucleon states 
$N_1$ (proton) and $N_2$, neglecting the energy difference between the scattered 
nucleon $N_2^\prime$ and the initial nucleon $N_2$.
$M_{J; {\bar p}p}(\mathbf{p}_1)$ ($M_{J_1; {\bar p}p}(\mathbf{p}_1)$) is the
invariant amplitude of the $\chi_J$ ($\chi_{J_1}$) production in the antiproton-nucleon
annihilation. $M_{\bar p N}(\mathbf{q}_2)$,
$M_{JN}(\mathbf{q}_2)$ and 
$M_{JN^\prime;J_1N}(\mathbf{q}_2)$ are, respectively, the invariant 
amplitudes of the antiproton and $\chi_J$ elastic scattering and of the nondiagonal transition
$\chi_{J_1} \to \chi_J$ on a nucleon.
The inverse propagators of the intermediate antiproton and charmonium states are
\begin{eqnarray}
   -D_{\bar p}(p_{\bar p}^\prime) &=& (p_{\bar p}^\prime)^2-m^2+i\varepsilon~, \label{D_barp}\\
   -D_J(k_1) &=& k_1^2-m_J^2+i\varepsilon~,  \label{D_J}
\end{eqnarray}
(and similar for $D_{J_1}(k_1)$) where $p_{\bar p}^\prime=p_{\bar p}+q_2$ 
and $k_1=p_{\bar p}+p_1$; 
$m$ and $m_J$ are the nucleon and charmonium masses, respectively.  
The normalization of the $\chi_J N$ elastic scattering amplitude is chosen such
that the optical theorem is 
\begin{equation}
   \mbox{Im}M_{J N}(0)=2p_{\rm lab}m\sigma_{JN}^{\rm tot}~, \label{optTheorem}
\end{equation}
and similar for the other elementary amplitudes. 
The differential cross section of the inclusive 
charmonium $\chi_J$ production on the nucleus $A$ is
\begin{eqnarray}
   d\sigma_{\bar p A \to \chi_J (A-1)^*} &=& \frac{2\pi\delta(E_{\bar p}+E_1-\omega)}{2p_{\rm lab}}
   \sum_{N_1} \sum_{\psi_{A-1}} |M^{J}(1)  \nonumber \\
       &&    +\sum_{N_2}(M^J(2,1)+M^J(1,2)+M^{J_1J}(1,2))|^2
   \frac{d^3k}{(2\pi)^32\omega}~,      \label{dsigma}
\end{eqnarray}
where the summations over all involved nucleon states ($N_1$, $N_2$)
and over all possible states of the final nucleus ($\psi_{A-1}$) are taken.
The many-body wave functions are normalized as
\begin{equation}
   \int d^3x_1 ...d^3x_A 
   |\psi_{A}(\mathbf{x}_1,\mathbf{x}_2,...,\mathbf{x}_A)|^2 = 1~.  \label{norma}
\end{equation} 
We have already implicitly applied the independent particle model for the nucleons
in the target nucleus by neglecting all position and momentum correlations between them
(including those due to antysymmetrization), i.e. we assumed that
\begin{equation}
   \psi_{A}(\mathbf{x}_1,\mathbf{x}_2,...,\mathbf{x}_A) 
    = \prod_{i=1}^A \phi_i(\mathbf{x}_i)    \label{indepPartModel}
\end{equation}
with $\phi_i(\mathbf{x}_i)$ being single-nucleon states normalized as
\begin{equation}
   \int d^3x_i |\phi_i(\mathbf{x}_i)|^2 = 1~.  \label{norma_i}
\end{equation}
This allowed us to separate-out the single particle energies $E_1$ and $E_2$ in 
Eqs. (\ref{M^J(1)})-(\ref{M^J1J(1,2)}). This is also the reason why the summation over
the first nucleon ($N_1$) is taken in Eq.(\ref{dsigma}) for the probabilities rather than
for the amplitudes.

The inverse propagators (\ref{D_barp}),(\ref{D_J}) can be simplified if one treats nucleons
nonrelativistically which gives
\begin{eqnarray}
   -D_{\bar p}(p_{\bar p}+q_2) &=& 2p_{\rm lab}(-q_2^z+i\varepsilon)~,\label{D_pbar_long} \\
   -D_J(p_{\bar p}+p_1) &=& 2p_{\rm lab}(\Delta_J^0-p_1^z+i\varepsilon)~, \label{D_J_long}
\end{eqnarray}
where
\begin{equation}
   \Delta_J^0=\frac{m^2+E_1^2+2E_{\bar p}E_1-m_J^2}{2p_{\rm lab}}~,    \label{DeltaJ^0}
\end{equation}
and $z$-axis is directed along the $\bar p$ beam momentum.
For the calculations of the production amplitudes on a nucleus we apply 
the GEA approach \cite{Frankfurt:1996xx,Sargsian01}. This approach 
is based on the coordinate representation of the propagator 
\begin{equation}
   \frac{1}{\Delta_{J}^0-p_1^z+i\varepsilon}
     =-i\int dz^0 \Theta(z^0) \mbox{e}^{i(\Delta_{J}^0-p_1^z)z^0}~,   \label{Dcoord}
\end{equation}
(and similar for the antiproton propagator)
and on the assumption that the elementary amplitudes depend on the transverse momentum transfer
only, i.e. $M_{J; {\bar p}p}(\mathbf{p}_{1t})$, $M_{J_1; {\bar p}p}(\mathbf{p}_{1t})$,
$M_{\bar p N}(\mathbf{t}_2)$, $M_{JN}(\mathbf{t}_2)$ and
$M_{J N^\prime; J_1 N}(\mathbf{t}_2)$.
Here $\mathbf{t}_2 \equiv \mathbf{q}_{2t} =\mathbf{k}_t-\mathbf{p}_{1t}$
is the transverse momentum transfer in the elastic scattering
$\bar p N \to \bar p N^\prime$, $\chi_{J} N \to \chi_{J} N^\prime$
or in the nondiagonal transition $\chi_{J_1} N \to \chi_{J} N^\prime$.
Then the amplitudes (\ref{M^J(2,1)}),(\ref{M^J(1,2)}) and (\ref{M^J1J(1,2)})
take the following form:
\begin{eqnarray}
   M^J(2,1) &=& 
   \frac{i}{\sqrt{2E_1} 2E_2 2p_{\rm lab} (2\pi)^2} \int d^3x_1 ...d^3x_A
      \psi^*_{A-1}(\mathbf{x}_2,\mathbf{x}_3,...,\mathbf{x}_A) \nonumber \\
   &\times& \psi_{A}(\mathbf{x}_1,\mathbf{x}_2,...,\mathbf{x}_A)
            \Theta(z_1-z_2) \mbox{e}^{-i(\mathbf{k}-\mathbf{p}_{\bar p})\mathbf{x}_1} \nonumber \\
   &\times& \int d^2t_2 \mbox{e}^{-i\mathbf{t}_2(\mathbf{b}_2-\mathbf{b}_1)}
            M_{J; {\bar p}p}(\mathbf{k}_t-\mathbf{t}_2) 
            M_{\bar p N}(\mathbf{t}_2)~,    \label{M^J_GEA(2,1)}\\
   M^J(1,2) &=& 
   \frac{i}{\sqrt{2E_1} 2E_2 2p_{\rm lab} (2\pi)^2} \int d^3x_1 ...d^3x_A
      \psi^*_{A-1}(\mathbf{x}_2,\mathbf{x}_3,...,\mathbf{x}_A) \nonumber \\
   &\times& \psi_{A}(\mathbf{x}_1,\mathbf{x}_2,...,\mathbf{x}_A)
            \Theta(z_2-z_1) \mbox{e}^{-i(\mathbf{k}-\mathbf{p}_{\bar p})\mathbf{x}_1
             +i(\Delta_{J}^0+p_{\rm lab}-k^z)(z_2-z_1)} \nonumber \\
   &\times& \int d^2t_2 \mbox{e}^{-i\mathbf{t}_2(\mathbf{b}_2-\mathbf{b}_1)}
            M_{J N}(\mathbf{t}_2) 
            M_{J; {\bar p}p}(\mathbf{k}_t-\mathbf{t}_2)~,    \label{M^J_GEA(1,2)}\\
   M^{J_1J}(1,2) &=& 
   \frac{i}{\sqrt{2E_1} 2E_2 2p_{\rm lab} (2\pi)^2} \int d^3x_1 ...d^3x_A
      \psi^*_{A-1}(\mathbf{x}_2,\mathbf{x}_3,...,\mathbf{x}_A) \nonumber \\
   &\times& \psi_{A}(\mathbf{x}_1,\mathbf{x}_2,...,\mathbf{x}_A)
            \Theta(z_2-z_1) \mbox{e}^{-i(\mathbf{k}-\mathbf{p}_{\bar p})\mathbf{x}_1
             +i(\Delta_{J_1}^0+p_{\rm lab}-k^z)(z_2-z_1)} \nonumber \\
   &\times& \int d^2t_2 \mbox{e}^{-i\mathbf{t}_2(\mathbf{b}_2-\mathbf{b}_1)}
            M_{J N^\prime; J_1 N}(\mathbf{t}_2) 
            M_{J_1; {\bar p}p}(\mathbf{k}_t-\mathbf{t}_2)~.    \label{M^J1J_GEA(1,2)}
\end{eqnarray}

The amplitudes squared and summed over all possible states of the final nucleus
$(A-1)$ can be evaluated by using the completeness relation:
\begin{eqnarray}
   && \sum_{\psi_{A-1}} \psi_{A-1}(\tilde\mathbf{x}_2,\tilde\mathbf{x}_3,\ldots,\tilde\mathbf{x}_A)    
   \psi^*_{A-1}(\mathbf{x}_2,\mathbf{x}_3,\ldots,\mathbf{x}_A)   \nonumber \\
   &&= \delta^{(3)}(\tilde\mathbf{x}_2-\mathbf{x}_2) \delta^{(3)}(\tilde\mathbf{x}_3-\mathbf{x}_3)  
\ldots \delta^{(3)}(\tilde\mathbf{x}_A-\mathbf{x}_A)~.    \label{CR}
\end{eqnarray}
For the impulse approximation (IA)-term we have
\begin{eqnarray}
   \sum_{\psi_{A-1}} |M^J(1)|^2
      &=& \frac{|M_{J; {\bar p}p}(\mathbf{k}_t)|^2}{2E_1}
    \int d^3\tilde\mathbf{x}_1 d^3\mathbf{x}_1 \cdots d^3\mathbf{x}_A
     \nonumber \\
   &&\times \psi_{A}^*(\tilde\mathbf{x}_1,\mathbf{x}_2,\ldots,\mathbf{x}_A)
    \psi_{A}(\mathbf{x}_1,\mathbf{x}_2,\ldots,\mathbf{x}_A)    
    \mbox{e}^{i(\mathbf{k}-\mathbf{p}_{\bar p})(\tilde\mathbf{x}_1-\mathbf{x}_1)} \nonumber \\
   &=& \frac{|M_{J; {\bar p}p}(\mathbf{k}_t)|^2}{2E_1}
       \int d^3X f_1(\mathbf{X},\mathbf{k}-\mathbf{p}_{\bar p})~,   \label{IAterm}
\end{eqnarray}
where 
\begin{equation}
   f_1(\mathbf{X},\mathbf{p})=\int d^3x \phi_1^*\left(\mathbf{X}+\frac{\mathbf{x}}{2}\right)
   \phi_1\left(\mathbf{X}-\frac{\mathbf{x}}{2}\right) 
   \mbox{e}^{i\mathbf{p}\mathbf{x}}              \label{WF}
\end{equation}
is the Wigner function (i.e. the phase space occupation number) of the struck nucleon.
To obtain the last form of Eq.(\ref{IAterm}), we directly applied the independent 
particle model relation (\ref{indepPartModel}) and introduced the new variables
$\mathbf{X}=(\tilde\mathbf{x}_1+\mathbf{x}_1)/2$ and 
$\mathbf{x}=\tilde\mathbf{x}_1-\mathbf{x}_1$.

The squares of the amplitudes of Figs.~\ref{fig:pbarChi_dia} (b,c,d) 
are calculated as
\begin{eqnarray}
\sum_{\psi_{A-1}} |M^J(2,1)|^2 &=& 
          \frac{1}{(2\pi)^4 2E_1 (4mp_{\rm lab})^2}
          \int d^3\tilde\mathbf{x}_1 d^3\mathbf{x}_1 \cdots d^3\mathbf{x}_A 
          \nonumber \\
   &&\times \psi_{A}^*(\tilde\mathbf{x}_1,\mathbf{x}_2,\ldots,\mathbf{x}_A)
     \psi_{A}(\mathbf{x}_1,\mathbf{x}_2,\ldots,\mathbf{x}_A)
     \mbox{e}^{ i\mathbf{k}_t(\tilde\mathbf{b}_1-\mathbf{b}_1)
               +i(k^z-p_{\rm lab})(\tilde{z}_1-z_1)}    \nonumber \\
   &&\times \Theta(\tilde{z}_1-z_2) \Theta(z_1-z_2) 
            \int d^2\tilde{t}_2 d^2t_2 
            \mbox{e}^{i\tilde\mathbf{t}_2(\mathbf{b}_2-\tilde\mathbf{b}_1)
                      -i\mathbf{t}_2(\mathbf{b}_2-\mathbf{b}_1)}  \nonumber \\
   &&\times M_{J; {\bar p}p}^*(\mathbf{k}_t-\tilde\mathbf{t}_2)
            M_{\bar p N}^*(\tilde\mathbf{t}_2) 
            M_{J; {\bar p}p}(\mathbf{k}_t-\mathbf{t}_2)
            M_{\bar p N}(\mathbf{t}_2)  \nonumber \\
&=& \frac{1}{(2\pi)^2 2E_1 (4mp_{\rm lab})^2}
       \int d^2t_2 |M_{J; {\bar p}p}(\mathbf{k}_t-\mathbf{t}_2)|^2 
                   |M_{\bar p N}(\mathbf{t}_2)|^2\,  \nonumber \\
   &&\times\int d^3X f_1(\mathbf{X},\mathbf{k}_t-\mathbf{t}_2,k^z-p_{\rm lab})
          \int\limits_{-\infty}^Z dz_2\, |\phi_2(\mathbf{B},z_2)|^2~,  \label{M^J(2,1)^2}\\
\sum_{\psi_{A-1}} |M^J(1,2)|^2 &=& 
          \frac{1}{(2\pi)^4 2E_1 (4mp_{\rm lab})^2}
          \int d^3\tilde\mathbf{x}_1 d^3\mathbf{x}_1 \cdots d^3\mathbf{x}_A 
          \nonumber \\
   &&\times \psi_{A}^*(\tilde\mathbf{x}_1,\mathbf{x}_2,\ldots,\mathbf{x}_A)
     \psi_{A}(\mathbf{x}_1,\mathbf{x}_2,\ldots,\mathbf{x}_A)
     \mbox{e}^{i\mathbf{k}_t(\tilde\mathbf{b}_1-\mathbf{b}_1)
               +i\Delta_{J}^0(\tilde{z}_1-z_1)}    \nonumber \\
   &&\times \Theta(z_2-\tilde{z}_1) \Theta(z_2-z_1) 
            \int d^2\tilde{t}_2 d^2t_2 
            \mbox{e}^{i\tilde\mathbf{t}_2(\mathbf{b}_2-\tilde\mathbf{b}_1)
                      -i\mathbf{t}_2(\mathbf{b}_2-\mathbf{b}_1)}  \nonumber \\
   &&\times M_{J N}^*(\tilde\mathbf{t}_2) 
            M_{J; {\bar p}p}^*(\mathbf{k}_t-\tilde\mathbf{t}_2)   
            M_{J N}(\mathbf{t}_2) 
            M_{J; {\bar p}p}(\mathbf{k}_t-\mathbf{t}_2) \nonumber \\
&=& \frac{1}{(2\pi)^2 2E_1 (4mp_{\rm lab})^2}
       \int d^2t_2 |M_{J N}(\mathbf{t}_2)|^2\,
                   |M_{J; {\bar p}p}(\mathbf{k}_t-\mathbf{t}_2)|^2 \nonumber \\
   &&\times\int d^3X f_1(\mathbf{X},\mathbf{k}_t-\mathbf{t}_2,\Delta_{J}^0)
          \int\limits_Z^{+\infty} dz_2\, |\phi_2(\mathbf{B},z_2)|^2~,  \label{M^J(1,2)^2}\\
\sum_{\psi_{A-1}} |M^{J_1J}(1,2)|^2 &=& 
   \frac{1}{(2\pi)^4 2E_1 (4mp_{\rm lab})^2}
   \int d^3\tilde\mathbf{x}_1 d^3\mathbf{x}_1 \cdots d^3\mathbf{x}_A 
     \nonumber \\
   &&\times \psi_{A}^*(\tilde\mathbf{x}_1,\mathbf{x}_2,\ldots,\mathbf{x}_A)
     \psi_{A}(\mathbf{x}_1,\mathbf{x}_2,\ldots,\mathbf{x}_A)
     \mbox{e}^{i\mathbf{k}_t(\tilde\mathbf{b}_1-\mathbf{b}_1)
               +i\Delta_{J_1}^0(\tilde{z}_1-z_1)}    \nonumber \\
   &&\times \Theta(z_2-\tilde{z}_1) \Theta(z_2-z_1) 
            \int d^2\tilde{t}_2 d^2t_2 
            \mbox{e}^{i\tilde\mathbf{t}_2(\mathbf{b}_2-\tilde\mathbf{b}_1)
                      -i\mathbf{t}_2(\mathbf{b}_2-\mathbf{b}_1)}  \nonumber \\
   &&\times M_{J N^\prime; J_1 N}^*(\tilde\mathbf{t}_2) 
            M_{J_1; {\bar p}p}^*(\mathbf{k}_t-\tilde\mathbf{t}_2)   
            M_{J N^\prime; J_1 N}(\mathbf{t}_2) 
            M_{J_1; {\bar p}p}(\mathbf{k}_t-\mathbf{t}_2) \nonumber \\
   &=& \frac{1}{(2\pi)^2 2E_1 (4mp_{\rm lab})^2}
       \int d^2t_2 |M_{J N^\prime; J_1 N}(\mathbf{t}_2)|^2\,
                   |M_{J_1; {\bar p}p}(\mathbf{k}_t-\mathbf{t}_2)|^2 \nonumber \\
  &&\times\int d^3X f_1(\mathbf{X},\mathbf{k}_t-\mathbf{t}_2,\Delta_{J_1}^0)
          \int\limits_Z^{+\infty} dz_2\, |\phi_2(\mathbf{B},z_2)|^2~.  \label{M^J1J(1,2)^2}
\end{eqnarray}
The last form of Eqs.(\ref{M^J(2,1)^2}),(\ref{M^J(1,2)^2}),(\ref{M^J1J(1,2)^2}) 
is obtained assuming that the momentum scale
of the elementary amplitudes variation is much larger than $1/L$, where $L \sim 1$ fm is the
characterisic scale on which the nucleon wave function changes. 
(If $|\mathbf{b}_1-\mathbf{b}_2| \sim L$ or $|\tilde\mathbf{b}_1-\mathbf{b}_2| \sim L$
then the exponent $\exp\{i\tilde\mathbf{t}_2(\mathbf{b}_2-\tilde\mathbf{b}_1)
-i\mathbf{t}_2(\mathbf{b}_2-\mathbf{b}_1)\}$ in the first Eqs.(\ref{M^J(2,1)^2}),(\ref{M^J(1,2)^2}),
(\ref{M^J1J(1,2)^2}) oscillates rapidly as a function 
of $\mathbf{t}_2$ or $\tilde\mathbf{t}_2$ and the integration over $d^2t_2d^2\tilde{t}_2$
gives almost zero.) 
This allows us to make the replacement $\phi_2(\mathbf{x}_2) \to \phi_2(\mathbf{B},z_2)$,
where $\mathbf{B}=(\tilde\mathbf{b}_1+\mathbf{b}_1)/2$ and perform the integration
over $d^2b_2$. 

Let us discuss now the interference terms. The leading ones are between
the IA-diagram (Fig.~\ref{fig:pbarChi_dia}a) and the elastic 
rescattering diagrams (Fig.~\ref{fig:pbarChi_dia}b,c):
\begin{eqnarray}
   && \sum_{\psi_{A-1}} M^J(2,1) M^{J*}(1)  +\mbox{c.c.} 
      = \frac{i M_{J; {\bar p}p}^*(\mathbf{k}_t)}{2E_1 (2\pi)^2 4mp_{\rm lab}} \nonumber \\
   &&\times \int d^3\tilde\mathbf{x}_1 d^3\mathbf{x}_1 \cdots d^3\mathbf{x}_A
     \psi_{A}^*(\tilde\mathbf{x}_1,\mathbf{x}_2,\ldots,\mathbf{x}_A) 
     \psi_{A}(\mathbf{x}_1,\mathbf{x}_2,\ldots,\mathbf{x}_A) \nonumber \\
   &&\times \mbox{e}^{i(\mathbf{k}-\mathbf{p}_{\bar p})(\tilde\mathbf{x}_1-\mathbf{x}_1)}
     \Theta(z_1-z_2)
     \int d^2t_2 
            \mbox{e}^{-i\mathbf{t}_2(\mathbf{b}_2-\mathbf{b}_1)}
             M_{J_; {\bar p}p}(\mathbf{k}_t-\mathbf{t}_2)
             M_{\bar p N}(\mathbf{t}_2) + \mbox{c.c.} \nonumber \\
   &=& \frac{iM_{\bar p N}(0)}{4mp_{\rm lab}}\frac{|M_{J; {\bar p}p}(\mathbf{k}_t)|^2}{2E_1} 
       \int d^3X  f_1(\mathbf{X},\mathbf{k}-\mathbf{p}_{\bar p})     
       \int\limits_{-\infty}^Z dz_2 |\phi_2(\mathbf{B},z_2)|^2
                                            + \mbox{c.c.}~, \label{M^J(2,1)M^J(1)} \\
   && \sum_{\psi_{A-1}} M^J(1,2) M^{J*}(1)  +\mbox{c.c.} 
      = \frac{i M_{J; {\bar p}p}^*(\mathbf{k}_t)}{2E_1 (2\pi)^2 4mp_{\rm lab}} \nonumber \\
   &&\times \int d^3\tilde\mathbf{x}_1 d^3\mathbf{x}_1 \cdots d^3\mathbf{x}_A
     \psi_{A}^*(\tilde\mathbf{x}_1,\mathbf{x}_2,\ldots,\mathbf{x}_A) 
     \psi_{A}(\mathbf{x}_1,\mathbf{x}_2,\ldots,\mathbf{x}_A) \nonumber \\
   &&\times \mbox{e}^{i(\mathbf{k}-\mathbf{p}_{\bar p})(\tilde\mathbf{x}_1-\mathbf{x}_1)
              +i(\Delta_{J}^0+p_{\rm lab}-k^z)(z_2-z_1)}
     \Theta(z_2-z_1) \nonumber \\
   &&\times \int d^2t_2 
            \mbox{e}^{-i\mathbf{t}_2(\mathbf{b}_2-\mathbf{b}_1)}
             M_{J N}(\mathbf{t}_2) 
             M_{J_; {\bar p}p}(\mathbf{k}_t-\mathbf{t}_2) + \mbox{c.c.} \nonumber \\
   &=& \frac{iM_{JN}(0)}{4mp_{\rm lab}}\frac{|M_{J; {\bar p}p}(\mathbf{k}_t)|^2}{2E_1} 
       \int d^3X  f_1\left(\mathbf{X},\mathbf{k}_t,\frac{k^z-p_{\rm lab}+\Delta^0_{J}}{2}\right)
                                                                           \nonumber \\     
   &&\times \int\limits_{Z}^{+\infty} dz_2 |\phi_2(\mathbf{B},z_2)|^2\,
            \mbox{e}^{i(\Delta_{J}^0-k^z+p_{\rm lab})(z_2-Z)}  + \mbox{c.c.}~, \label{M^J(1,2)M^J(1)}
\end{eqnarray}
where we again assumed the smallness of the matrix element variation on the momentum scale 
of the order of $L^{-1}$. By using the optical theorem (\ref{optTheorem})
we see that the both interference terms (\ref{M^J(2,1)M^J(1)}) and (\ref{M^J(1,2)M^J(1)}) are the
absorptive corrections to the IA-term (\ref{IAterm}). (For the term (\ref{M^J(1,2)M^J(1)})
one has to require in addition  that $k^z=p_{\rm lab}+\Delta_{J}^0$, i.e. restrict the kinematics
of the final charmonium $\chi_J$ to the quasifree regime, see also Eq.(\ref{kin_constr}).)
On the other hand, the interference term between the IA-diagram (Fig.~\ref{fig:pbarChi_dia}a) and 
the nondiagonal transition diagram (Fig.~\ref{fig:pbarChi_dia}d) has a pure quantum mechanical
origin and can not be interpreted in a probabilistic picture:
\begin{eqnarray}
   &&\sum_{\psi_{A-1}} M^{J_1J}(1,2) M^{J*}(1)  + \mbox{c.c.}
      = \frac{i M_{J; {\bar p}p}^*(\mathbf{k}_t)}{2E_1 (2\pi)^2 4mp_{\rm lab}} \nonumber \\
   &&\times \int d^3\tilde\mathbf{x}_1 d^3\mathbf{x}_1 \cdots d^3\mathbf{x}_A
     \psi_{A}^*(\tilde\mathbf{x}_1,\mathbf{x}_2,\ldots,\mathbf{x}_A) 
     \psi_{A}(\mathbf{x}_1,\mathbf{x}_2,\ldots,\mathbf{x}_A) \nonumber \\
   &&\times \mbox{e}^{i(\mathbf{k}-\mathbf{p}_{\bar p})(\tilde\mathbf{x}_1-\mathbf{x}_1)
              +i(\Delta_{J_1}^0+p_{\rm lab}-k^z)(z_2-z_1)}
     \Theta(z_2-z_1) \nonumber \\
   &&\times \int d^2t_2 
            \mbox{e}^{-i\mathbf{t}_2(\mathbf{b}_2-\mathbf{b}_1)}
             M_{J N^\prime; J_1 N}(\mathbf{t}_2) 
             M_{J_1; {\bar p}p}(\mathbf{k}_t-\mathbf{t}_2) + \mbox{c.c.} \nonumber \\
   &=& \frac{i M_{J; {\bar p}p}^*(\mathbf{k}_t)}{2E_1 4mp_{\rm lab}} 
        M_{J N^\prime; J_1 N}(0) M_{J_1; {\bar p}p}(\mathbf{k}_t)
       \int d^3X  f_1\left(\mathbf{X},\mathbf{k}_t,\frac{k^z-p_{\rm lab}+\Delta^0_{J_1}}{2}\right)
                                                                           \nonumber \\     
   &&\times \int\limits_{Z}^{+\infty} dz_2 |\phi_2(\mathbf{B},z_2)|^2\,
            \mbox{e}^{i(\Delta_{J_1}^0-k^z+p_{\rm lab})(z_2-Z)}  + \mbox{c.c.}~. \label{M^J1J(1,2)M^J(1)}
\end{eqnarray}
Finally, the interference term between the charmonium elastic rescattering diagram 
(Fig.~\ref{fig:pbarChi_dia}c)
and the nondiagonal transition diagram (Fig.~\ref{fig:pbarChi_dia}d) is calculated
as follows:
\begin{eqnarray}
   &&\sum_{\psi_{A-1}} M^{J_1J}(1,2) M^{J*}(1,2) + \mbox{c.c.}
    = \frac{1}{(2\pi)^4 2E_1 (4mp_{\rm lab})^2} \nonumber \\
   &&\times \int d^3\tilde\mathbf{x}_1 d^3\mathbf{x}_1 \cdots d^3\mathbf{x}_A 
      \psi_{A}^*(\tilde\mathbf{x}_1,\mathbf{x}_2,\ldots,\mathbf{x}_A)
      \psi_{A}(\mathbf{x}_1,\mathbf{x}_2,\ldots,\mathbf{x}_A) \nonumber \\
   &&\times  \mbox{e}^{ i\mathbf{k}_t(\tilde\mathbf{b}_1-\mathbf{b}_1)
                       +i(\Delta_{J_1}^0-\Delta_{J}^0)z_2
                       +i\Delta_{J}^0\tilde{z}_1 - i\Delta_{J_1}^0z_1}
            \Theta(z_2-\tilde{z}_1) \Theta(z_2-z_1)     \nonumber \\
   &&\times \int d^2\tilde{t}_2 d^2t_2 
            \mbox{e}^{i\tilde\mathbf{t}_2(\mathbf{b}_2-\tilde\mathbf{b}_1)
                      -i\mathbf{t}_2(\mathbf{b}_2-\mathbf{b}_1)}
            M_{J N^\prime; J_1 N}(\mathbf{t}_2) 
            M_{J_1; {\bar p}p}(\mathbf{k}_t-\mathbf{t}_2)   
            M_{J N}^*(\tilde\mathbf{t}_2) 
            M_{J; {\bar p}p}^*(\mathbf{k}_t-\tilde\mathbf{t}_2)  + \mbox{c.c.}  \nonumber \\
   &=& \frac{1}{(2\pi)^2 2E_1 (4mp_{\rm lab})^2}
       \int d^2t_2 M_{J N^\prime; J_1 N}(\mathbf{t}_2) 
                   M_{J_1; {\bar p}p}(\mathbf{k}_t-\mathbf{t}_2)
                   M_{J N}^*(\mathbf{t}_2) M_{J; {\bar p}p}^*(\mathbf{k}_t-\mathbf{t}_2) \nonumber \\
   &&\times  \int d^3X 
     f_1\left(\mathbf{X},\mathbf{k}_t-\mathbf{t}_2,\frac{\Delta_{J_1}^0+\Delta_{J}^0}{2}\right)
    \int\limits_Z^{+\infty} dz_2\, 
     \mbox{e}^{i(\Delta_{J_1}^0-\Delta_{J}^0)(z_2-Z)}
     |\phi_2(\mathbf{B},z_2)|^2  + \mbox{c.c.}~.  \label{M^J1J(1,2)M^J(1,2)}
\end{eqnarray}
The interference terms between the antiproton rescattering diagram (Fig.~\ref{fig:pbarChi_dia}b)
and the charmonium rescattering and nondiagonal transition diagrams (Fig.~\ref{fig:pbarChi_dia}c,d)
disappear in our approximation since they include
the products of the factors $\Theta(z_2-z_1)\Theta(\tilde z_1 -z_2)$.

\subsection{Absorptive corrections}

The above formulas for the products of matrix elements can be generalized 
to take into account the multiple elastic rescattering effects 
(see Appendix \ref{MultScatt}).
The sum of the interference terms between the diagonal amplitudes 
(\ref{M^Jmult}) with elastic rescatterings of the antiproton and
$\chi_J$-charmonium on all possible nonoverlapping sets of nucleons
can be expressed as
\begin{eqnarray}
   && \sum_{\mbox{set1} \neq \mbox{set2}} \sum_{\psi_{A-1}} 
      M^J(1,\mbox{set1}) M^{J*}(1,\mbox{set2}) = 
      \frac{|M_{J; {\bar p}p}(\mathbf{k}_t)|^2}{2E_1}  
      \int d^3\tilde\mathbf{x}_1 d^3\mathbf{x}_1 \cdots d^3\mathbf{x}_A \nonumber \\
   &&\times \psi_{A}^*(\tilde\mathbf{x}_1,\mathbf{x}_2,...,\mathbf{x}_A)
            \psi_{A}(\mathbf{x}_1,\mathbf{x}_2,...,\mathbf{x}_A)
            \mbox{e}^{ i\Delta_J^0(\tilde z_1-z_1)
                      +i\mathbf{k}_t(\tilde\mathbf{b}_1-\mathbf{b}_1)} \nonumber \\
   &&\times \prod_{i=2}^A\left(1 
                                 + \frac{i}{4mp_{\rm lab}} \left[
                  M_{\bar p N}(0) \Theta(z_1-z_i) \delta^{(2)}(\mathbf{b}_i-\mathbf{b}_1)
                + M_{JN}(0) \Theta(z_i-z_1) \delta^{(2)}(\mathbf{b}_i-\mathbf{b}_1) 
                                                           \right.\right. \nonumber \\
   &&\hspace{3cm} \left.\left. 
                - M^*_{\bar p N}(0) \Theta(\tilde z_1-z_i) \delta^{(2)}(\mathbf{b}_i-\tilde\mathbf{b}_1)
                - M^*_{JN}(0) \Theta(z_i-\tilde z_1) \delta^{(2)}(\mathbf{b}_i-\tilde\mathbf{b}_1) 
                          \right]\right)  \nonumber \\
   &=& \frac{|M_{J; {\bar p}p}(\mathbf{k}_t)|^2}{2E_1}
       \int d^3X f_1(\mathbf{X},\mathbf{k}_t,\Delta_J^0) \nonumber \\
   &&\times \prod_{i=2}^A\left(1 - \sigma_{\bar p N}^{\rm tot} \int\limits_{-\infty}^Z dz_i
                      |\phi_i(\mathbf{B},z_i)|^2
                          -\sigma_{J N}^{\rm tot} \int\limits_Z^{+\infty} dz_i
                      |\phi_i(\mathbf{B},z_i)|^2\right)~,   \label{DirectTerm}
\end{eqnarray}
where we again assumed the slowness of the ground state wave function variation
with transverse coordinates. The sets of nucleons-scatterers are denoted as
"set1" and "set2". We neglect in Eq. (\ref{DirectTerm}) the product
terms with the same nucleon-scatterer in the direct and conjugated amplitudes which
give the proper rescattering contributions discussed in the next
subsection. Note that the struck nucleon $N_1$ is fixed in the both 
amplitudes and is excluded from the sets of scatterers.
In Eq.(\ref{DirectTerm}) we assumed that the motion of nucleons 
inside the nucleus is quasiclassical, i.e. the product 
$\phi_1^*(\mathbf{X}+\mathbf{x}/2)\phi_1(\mathbf{X}-\mathbf{x}/2)$
in the Wigner function (\ref{WF}) changes much faster as a function 
of the relative coordinate $x$ than as a function
of the center-of-mass variable $X$. This allows to replace $\mathbf{x}_1 \to \mathbf{X}$
and $\tilde\mathbf{x}_1 \to \mathbf{X}$ in the multiple product factors
and perform the integration of the wave functions over the relative coordinate
$\mathbf{x}$ separately.

In the case of identical nucleons and large $A$ the multiple product factors
are reduced to the exponential absorption for the antiproton
and charmonium
\begin{eqnarray}
      && \left(1-\sigma_{\bar p N}^{\rm tot}\int\limits_{-\infty}^Z dz_2
                                         |\phi_2(\mathbf{B},z_2)|^2
             -\sigma_{J N}^{\rm tot} \int\limits_Z^{+\infty} dz_2
                                         |\phi_2(\mathbf{B},z_2)|^2\right)^{A-1} \nonumber \\
      && \simeq \exp\left(-\sigma_{\bar p N}^{\rm tot}
                   \int\limits_{-\infty}^Z dz_2 \rho(\mathbf{B},z_2)
                      -\sigma_{J N}^{\rm tot}  \int\limits_Z^{+\infty} dz_2
                                                \rho(\mathbf{B},z_2)
                    \right)~.    \label{ExpAbs}
\end{eqnarray}
Here $\rho(\mathbf{B},z_2)=A|\phi_2(\mathbf{B},z_2)|^2$ is the nucleon density. 
Thus, Eq.(\ref{DirectTerm})
is an extension of the IA-term (\ref{IAterm}) for the absorption of the
incoming antiproton and of the outgoing $\chi_J$-charmonium.

The leading order contribution of the nondiagonal transition
(see Fig.~\ref{fig:pbarChi_nondiag_mult} in Appendix \ref{MultScatt}) 
to the total amplitude squared appears as the interference
of the diagonal (\ref{M^Jmult}) and the nondiagonal (\ref{M^J1Jmult}) 
amplitudes summed over all possible nonoverlapping sets of nucleons-scatterers:
\begin{eqnarray}
   && \sum_{\mbox{set1} \neq \mbox{set2}} \sum_{\psi_{A-1}} 
      M^{J_1J}(1,2,\mbox{set1}) M^{J*}(1,\mbox{set2}) + \mbox{c.c.} 
      = \frac{iM_{J; {\bar p}p}^*(\mathbf{k}_t)}{2E_1 4mp_{\rm lab}} \nonumber \\
   &&\times M_{J N^\prime; J_1 N}(0) M_{J_1; {\bar p}p}(\mathbf{k}_t)
           \int d^3\tilde\mathbf{x}_1 d^3\mathbf{x}_1 \cdots d^3\mathbf{x}_A
     \psi_{A}^*(\tilde\mathbf{x}_1,\mathbf{x}_2,\ldots,\mathbf{x}_A) 
     \psi_{A}(\mathbf{x}_1,\mathbf{x}_2,\ldots,\mathbf{x}_A) \nonumber \\
   &&\times \Theta(z_2-z_1) \delta^{(2)}(\mathbf{b}_2-\mathbf{b}_1)
     \mbox{e}^{ i\mathbf{k}_t(\tilde\mathbf{b}_1-\mathbf{b}_1)
               +i\Delta_J^0 \tilde{z}_1 - i\Delta_{J_1}^0 z_1 
               +i(\Delta_{J_1}^0-\Delta_J^0) z_2
              }        \nonumber \\
   &&\times \prod_{i=3}^A\left(1 + \frac{i}{4mp_{\rm lab}}\left[
       M_{\bar p N}(0) \Theta(z_1-z_i) \delta^{(2)}(\mathbf{b}_i-\mathbf{b}_1)
                                              \right.\right. \nonumber \\
   &&\hspace{3.5cm}
       +M_{J_1 N}(0) \Theta(z_i-z_1) \Theta(z_2-z_i) \delta^{(2)}(\mathbf{b}_i-\mathbf{b}_1)
       +M_{J N}(0) \Theta(z_i-z_2) \delta^{(2)}(\mathbf{b}_i-\mathbf{b}_1)   \nonumber \\ 
   &&\hspace{3.3cm}   \left.\left.
       -M^*_{\bar p N}(0) \Theta(\tilde z_1-z_i) \delta^{(2)}(\mathbf{b}_i-\tilde\mathbf{b}_1)
       -M^*_{JN}(0) \Theta(z_i-\tilde z_1) \delta^{(2)}(\mathbf{b}_i-\tilde\mathbf{b}_1) 
                           \right] \right) + \mbox{c.c.}  \nonumber \\
   &=& \frac{iM_{J; {\bar p}p}^*(\mathbf{k}_t)}{2E_1 4mp_{\rm lab}}
        M_{J N^\prime; J_1 N}(0) M_{J_1; {\bar p}p}(\mathbf{k}_t)
        \int d^3X 
        f_1\left(\mathbf{X},\mathbf{k}_t,\frac{\Delta_J^0+\Delta_{J_1}^0}{2}\right)
            \int\limits_Z^{+\infty} dz_2\, |\phi_2(\mathbf{B},z_2)|^2  
             \mbox{e}^{i(\Delta_{J_1}^0-\Delta^0_J)(z_2-Z)}       \nonumber \\   
   &&\times   \prod_{i=3}^A \left(1    
                    - \sigma_{\bar p N}^{\rm tot}\int\limits_{-\infty}^Z dz_i
                      |\phi_i(\mathbf{B},z_i)|^2
                    + \frac{i[M_{J_1 N}(0)-M^*_{JN}(0)]}{4mp_{\rm lab}}
                      \int\limits_{Z}^{z_2} dz_i 
                      |\phi_i(\mathbf{B},z_i)|^2     \right. \nonumber \\
   && \hspace{2cm} \left.                 
                    - \sigma_{J N}^{\rm tot} 
                      \int\limits_{z_2}^{+\infty} dz_i
                      |\phi_i(\mathbf{B},z_i)|^2 \right)  + \mbox{c.c.}~.   \label{InterfTerm}
\end{eqnarray}
The struck nucleon ($N_1$) and the nucleon on which the nondiagonal transition happen
($N_2$) are excluded from both sets of nucleon-scatterers. 
Without taking into account the absorptive correction, this expression is reduced 
to the interference term (\ref{M^J1J(1,2)M^J(1)}).

\subsection{Rescattering contributions}

We will now take into account the interference between the amplitudes 
where the elastic or nondiagonal rescattering happen on the same nucleon ($N_2$).
Fixing the struck nucleon ($N_1$) and the nucleon-scatterer ($N_2$) in the direct
and conjugated amplitudes we sum all possible interference terms with nonoverlapping sets
of other participating nucleons.

Four terms appear as the result.
(i) The term due to the antiproton elastic rescattering (c.f. Eq.(\ref{M^J(2,1)^2}))
given by the product of the direct and conjugated amplitudes (\ref{M^Jmult}):
\begin{eqnarray}
   &&  \sum_{\mbox{set1} \neq \mbox{set2}} \sum_{\psi_{A-1}} 
            M^J(2,1,\mbox{set1}) M^{J*}(2,1,\mbox{set2}) = 
            \frac{1}{(2\pi)^4 2E_1 (4mp_{\rm lab})^2}   \nonumber \\
   &&\times \int d^3\tilde\mathbf{x}_1 d^3\mathbf{x}_1 \cdots d^3\mathbf{x}_A                                
       \psi_{A}^*(\tilde\mathbf{x}_1,\mathbf{x}_2,\ldots,\mathbf{x}_A)
       \psi_{A}(\mathbf{x}_1,\mathbf{x}_2,\ldots,\mathbf{x}_A)
     \mbox{e}^{i\mathbf{k}_t(\tilde\mathbf{b}_1-\mathbf{b}_1)
               +i\Delta_{J}^0(\tilde{z}_1-z_1)}    \nonumber \\
   &&\times \Theta(\tilde{z}_1-z_2) \Theta(z_1-z_2) 
            \int d^2\tilde{t}_2 d^2t_2 
            \mbox{e}^{i\tilde\mathbf{t}_2(\mathbf{b}_2-\tilde\mathbf{b}_1)
                      -i\mathbf{t}_2(\mathbf{b}_2-\mathbf{b}_1)}
             M_{J; {\bar p}p}^*(\mathbf{k}_t-\tilde\mathbf{t}_2)           
             M_{\bar p N}^*(\tilde\mathbf{t}_2)  \nonumber \\  
   &&\times  M_{J; {\bar p}p}(\mathbf{k}_t-\mathbf{t}_2) 
             M_{\bar p N}(\mathbf{t}_2)            \nonumber \\ 
   &&\times         \prod_{i=3}^A\left(1 + \frac{i}{4mp_{\rm lab}}\left[
             M_{\bar p N}(0) \Theta(z_1-z_i) \delta^{(2)}(\mathbf{b}_i-\mathbf{b}_1)
            +M_{JN}(0)\Theta(z_i-z_1) \delta^{(2)}(\mathbf{b}_i-\mathbf{b}_1)    \right.\right. \nonumber \\
   &&\hspace{3.2cm}  \left.\left.  
            -M^*_{\bar p N}(0) \Theta(\tilde z_1-z_i) \delta^{(2)}(\mathbf{b}_i-\tilde\mathbf{b}_1)
            -M^*_{JN}(0) \Theta(z_i-\tilde z_1) \delta^{(2)}(\mathbf{b}_i-\tilde\mathbf{b}_1)
                         \right]\right)  \nonumber \\   
   &=&  \frac{1}{(2\pi)^2 2E_1 (4mp_{\rm lab})^2}
        \int d^2t_2 |M_{J; {\bar p}p}(\mathbf{k}_t-\mathbf{t}_2)|^2\,
                    |M_{\bar p N}(\mathbf{t}_2)|^2
              \int d^3X f_1(\mathbf{X},\mathbf{k}_t-\mathbf{t}_2,\Delta_{J}^0) \nonumber \\
   &&\times   \int\limits_{-\infty}^Z dz_2\, |\phi_2(\mathbf{B},z_2)|^2 
     \prod_{i=3}^A\left(1 - \sigma_{\bar p N}^{\rm tot} \int\limits_{-\infty}^Z dz_i
                      |\phi_i(\mathbf{B},z_i)|^2
                          -\sigma_{J N}^{\rm tot} \int\limits_Z^{+\infty} dz_i
                      |\phi_i(\mathbf{B},z_i)|^2\right)~.  \label{PbarRescTerm}
\end{eqnarray}

(ii) The diagonal term with rescattering (c.f. Eq.(\ref{M^J(1,2)^2}))
due to the product of the direct and conjugated amplitudes (\ref{M^Jmult}):
\begin{eqnarray}
   &&  \sum_{\mbox{set1} \neq \mbox{set2}} \sum_{\psi_{A-1}} 
            M^J(1,2,\mbox{set1}) M^{J*}(1,2,\mbox{set2}) = 
            \frac{1}{(2\pi)^4 2E_1 (4mp_{\rm lab})^2}   \nonumber \\
   &&\times \int d^3\tilde\mathbf{x}_1 d^3\mathbf{x}_1 \cdots d^3\mathbf{x}_A                                
       \psi_{A}^*(\tilde\mathbf{x}_1,\mathbf{x}_2,\ldots,\mathbf{x}_A)
       \psi_{A}(\mathbf{x}_1,\mathbf{x}_2,\ldots,\mathbf{x}_A)
     \mbox{e}^{i\mathbf{k}_t(\tilde\mathbf{b}_1-\mathbf{b}_1)
               +i\Delta_{J}^0(\tilde{z}_1-z_1)}    \nonumber \\
   &&\times \Theta(z_2-\tilde{z}_1) \Theta(z_2-z_1) 
            \int d^2\tilde{t}_2 d^2t_2 
            \mbox{e}^{i\tilde\mathbf{t}_2(\mathbf{b}_2-\tilde\mathbf{b}_1)
                      -i\mathbf{t}_2(\mathbf{b}_2-\mathbf{b}_1)}
            M_{J N}^*(\tilde\mathbf{t}_2) 
            M_{J; {\bar p}p}^*(\mathbf{k}_t-\tilde\mathbf{t}_2) \nonumber \\  
   &&\times M_{J N}(\mathbf{t}_2) 
            M_{J; {\bar p}p}(\mathbf{k}_t-\mathbf{t}_2)  \nonumber \\ 
   &&\times         \prod_{i=3}^A\left(1  + \frac{i}{4mp_{\rm lab}}\left[
            M_{\bar p N}(0) \Theta(z_1-z_i) \delta^{(2)}(\mathbf{b}_i-\mathbf{b}_1)
           +M_{JN}(0) \Theta(z_i-z_1) \delta^{(2)}(\mathbf{b}_i-\mathbf{b}_1) \right.\right. \nonumber \\
   &&\hspace{3cm}  \left.\left.  
           -M^*_{\bar p N}(0) \Theta(\tilde z_1-z_i) \delta^{(2)}(\mathbf{b}_i-\tilde\mathbf{b}_1)
           -M^*_{JN}(0)\Theta(z_i-\tilde z_1) \delta^{(2)}(\mathbf{b}_i-\tilde\mathbf{b}_1)
                         \right]\right)  \nonumber \\   
   &=&  \frac{1}{(2\pi)^2 2E_1 (4mp_{\rm lab})^2}
        \int d^2t_2 |M_{J N}(\mathbf{t}_2)|^2\,
                   |M_{J; {\bar p}p}(\mathbf{k}_t-\mathbf{t}_2)|^2
          \int d^3X f_1(\mathbf{X},\mathbf{k}_t-\mathbf{t}_2,\Delta_{J}^0) \nonumber \\
   &&\times \int\limits_Z^{+\infty} dz_2\, |\phi_2(\mathbf{B},z_2)|^2 
     \prod_{i=3}^A\left(1 - \sigma_{\bar p N}^{\rm tot} \int\limits_{-\infty}^Z dz_i
                      |\phi_i(\mathbf{B},z_i)|^2
                          -\sigma_{J N}^{\rm tot} \int\limits_Z^{+\infty} dz_i
                      |\phi_i(\mathbf{B},z_i)|^2\right)~.  \label{DiagRescTerm}
\end{eqnarray}

(iii) The nondiagonal rescattering term (c.f. Eq.(\ref{M^J1J(1,2)^2}))
due to the the product of the direct and conjugated amplitudes 
(\ref{M^J1Jmult}):
\begin{eqnarray}
   &&  \sum_{\mbox{set1} \neq \mbox{set2}} \sum_{\psi_{A-1}} 
            M^{J_1J}(1,2,\mbox{set1}) M^{J_1J*}(1,2,\mbox{set2}) = 
            \frac{1}{(2\pi)^4 2E_1 (4mp_{\rm lab})^2}   \nonumber \\
   &&\times \int d^3\tilde\mathbf{x}_1 d^3\mathbf{x}_1 \cdots d^3\mathbf{x}_A                                
       \psi_{A}^*(\tilde\mathbf{x}_1,\mathbf{x}_2,\ldots,\mathbf{x}_A)
       \psi_{A}(\mathbf{x}_1,\mathbf{x}_2,\ldots,\mathbf{x}_A)
       \mbox{e}^{i\mathbf{k}_t(\tilde\mathbf{b}_1-\mathbf{b}_1)
                 +i\Delta_{J_1}^0(\tilde{z}_1-z_1)}    \nonumber \\
   &&\times \Theta(z_2-\tilde{z}_1) \Theta(z_2-z_1) 
            \int d^2\tilde{t}_2 d^2t_2 
            \mbox{e}^{i\tilde\mathbf{t}_2(\mathbf{b}_2-\tilde\mathbf{b}_1)
                      -i\mathbf{t}_2(\mathbf{b}_2-\mathbf{b}_1)}  \nonumber \\
   &&\times M_{J N^\prime; J_1 N}^*(\tilde\mathbf{t}_2) 
            M_{J_1; {\bar p}p}^*(\mathbf{k}_t-\tilde\mathbf{t}_2)   
            M_{J N^\prime; J_1 N}(\mathbf{t}_2) 
            M_{J_1; {\bar p}p}(\mathbf{k}_t-\mathbf{t}_2) \nonumber \\
   &&\times         \prod_{i=3}^A\left(1 + \frac{i}{4mp_{\rm lab}}\left[
               M_{\bar p N}(0) \Theta(z_1-z_i) \delta^{(2)}(\mathbf{b}_i-\mathbf{b}_1)
                                              \right.\right. \nonumber \\
   &&\hspace{3cm}
             + M_{J_1 N}(0) \Theta(z_i-z_1)\Theta(z_2-z_i) \delta^{(2)}(\mathbf{b}_i-\mathbf{b}_1)
             + M_{JN}(0) \Theta(z_i-z_2) \delta^{(2)}(\mathbf{b}_i-\mathbf{b}_1)   \nonumber \\ 
   &&\hspace{3cm}  
             - M^*_{\bar p N}(0) \Theta(\tilde z_1-z_i) \delta^{(2)}(\mathbf{b}_i-\tilde\mathbf{b}_1) \nonumber \\
   &&\hspace{3cm} \left.\left.
             - M^*_{J_1 N}(0) \Theta(z_i-\tilde z_1) \Theta(z_2-z_i) \delta^{(2)}(\mathbf{b}_i-\tilde\mathbf{b}_1)
             - M^*_{JN}(0) \Theta(z_i-z_2) \delta^{(2)}(\mathbf{b}_i-\tilde\mathbf{b}_1)
                         \right] \right)  \nonumber \\     
   &=& \frac{1}{(2\pi)^2 2E_1 (4mp_{\rm lab})^2}
       \int d^2t_2 |M_{J N^\prime; J_1 N}(\mathbf{t}_2)|^2\,
                   |M_{J_1; {\bar p}p}(\mathbf{k}_t-\mathbf{t}_2)|^2 \nonumber \\
  &&\times\int d^3X f_1(\mathbf{X},\mathbf{k}_t-\mathbf{t}_2,\Delta_{J_1}^0)
          \int\limits_Z^{+\infty} dz_2\, |\phi_2(\mathbf{B},z_2)|^2   \nonumber \\
  &&\times \prod_{i=3}^A \left(1 - \sigma_{\bar p N}^{\rm tot}\int\limits_{-\infty}^Z dz_i
                                            |\phi_i(\mathbf{B},z_i)|^2 
                                 - \sigma_{J_1 N}^{\rm tot} \int\limits_{Z}^{z_2} dz_i 
                                            |\phi_i(\mathbf{B},z_i)|^2
                                 - \sigma_{J N}^{\rm tot} \int\limits_{z_2}^{+\infty} dz_i
                                            |\phi_i(\mathbf{B},z_i)|^2\right)~.   \label{NondiagRescTerm}
\end{eqnarray}

And, (iv) the interference of the nondiagonal and diagonal terms with rescattering
(c.f. Eq.(\ref{M^J1J(1,2)M^J(1,2)})) given by the product of the direct and conjugated
amplitudes (\ref{M^J1Jmult}),(\ref{M^Jmult}):
\begin{eqnarray}
   &&  \sum_{\mbox{set1} \neq \mbox{set2}} \sum_{\psi_{A-1}} 
            M^{J_1J}(1,2,\mbox{set1}) M^{J*}(1,2,\mbox{set2}) + \mbox{c.c.} = 
            \frac{1}{(2\pi)^4 2E_1 (4mp_{\rm lab})^2}   \nonumber \\
   &&\times \int d^3\tilde\mathbf{x}_1 d^3\mathbf{x}_1 \cdots d^3\mathbf{x}_A                                
       \psi_{A}^*(\tilde\mathbf{x}_1,\mathbf{x}_2,\ldots,\mathbf{x}_A)
       \psi_{A}(\mathbf{x}_1,\mathbf{x}_2,\ldots,\mathbf{x}_A) \nonumber \\
   &&\times  \mbox{e}^{ i\mathbf{k}_t(\tilde\mathbf{b}_1-\mathbf{b}_1)
                       +i(\Delta_{J_1}^0-\Delta_{J}^0)z_2
                       +i\Delta_{J}^0\tilde{z}_1 - i\Delta_{J_1}^0z_1}
            \Theta(z_2-\tilde{z}_1) \Theta(z_2-z_1)     \nonumber \\
   &&\times \int d^2\tilde{t}_2 d^2t_2 
            \mbox{e}^{i\tilde\mathbf{t}_2(\mathbf{b}_2-\tilde\mathbf{b}_1)
                      -i\mathbf{t}_2(\mathbf{b}_2-\mathbf{b}_1)}
            M_{J N^\prime; J_1 N}(\mathbf{t}_2) 
            M_{J_1; {\bar p}p}(\mathbf{k}_t-\mathbf{t}_2)   
            M_{J N}^*(\tilde\mathbf{t}_2) 
            M_{J; {\bar p}p}^*(\mathbf{k}_t-\tilde\mathbf{t}_2)  \nonumber \\
   &&\times         \prod_{i=3}^A\left(1 + \frac{i}{4mp_{\rm lab}}\left[
                  M_{\bar p N}(0) \Theta(z_1-z_i) \delta^{(2)}(\mathbf{b}_i-\mathbf{b}_1)
                + M_{J_1 N}(0) \Theta(z_i-z_1)\Theta(z_2-z_i) \delta^{(2)}(\mathbf{b}_i-\mathbf{b}_1)
                   \right.\right.               \nonumber \\
   &&\hspace{3cm} 
                + M_{JN}(0) \Theta(z_i-z_2) \delta^{(2)}(\mathbf{b}_i-\mathbf{b}_1)
                - M^*_{\bar p N}(0) \Theta(\tilde z_1-z_i) 
                  \delta^{(2)}(\mathbf{b}_i-\tilde\mathbf{b}_1) \nonumber \\
   &&\hspace{3cm} \left.\left.
                - M^*_{JN}(0) \Theta(z_i-\tilde z_1) \Theta(z_2-z_i)
                  \delta^{(2)}(\mathbf{b}_i-\tilde\mathbf{b}_1)
                - M^*_{JN}(0) \Theta(z_i-z_2)
                  \delta^{(2)}(\mathbf{b}_i-\tilde\mathbf{b}_1)
                         \right]\right)  + \mbox{c.c.}  \nonumber \\
   &=& \frac{1}{(2\pi)^2 2E_1 (4mp_{\rm lab})^2}
       \int d^2t_2 M_{J N^\prime; J_1 N}(\mathbf{t}_2) 
                   M_{J_1; {\bar p}p}(\mathbf{k}_t-\mathbf{t}_2)
                   M_{J N}^*(\mathbf{t}_2)
                   M_{J; {\bar p}p}^*(\mathbf{k}_t-\mathbf{t}_2)  \nonumber \\
   &&\times  \int d^3X 
     f_1\left(\mathbf{X},\mathbf{k}_t-\mathbf{t}_2,\frac{\Delta_{J_1}^0+\Delta_{J}^0}{2}\right)
     \int\limits_Z^{+\infty} dz_2\, 
     \mbox{e}^{i(\Delta_{J_1}^0-\Delta_{J}^0)(z_2-Z)}
     |\phi_2(\mathbf{B},z_2)|^2  \nonumber \\ 
   &&\times  \prod_{i=3}^A \left(1    
                    - \sigma_{\bar p N}^{\rm tot}\int\limits_{-\infty}^Z dz_i
                      |\phi_i(\mathbf{B},z_i)|^2     
                    + \frac{i[M_{J_1 N}(0)-M^*_{JN}(0)]}{4mp_{\rm lab}}
                      \int\limits_{Z}^{z_2} dz_i 
                      |\phi_i(\mathbf{B},z_i)|^2     \right.       \nonumber \\
   &&\hspace{1.5cm}   \left.
                    - \sigma_{J N}^{\rm tot} 
                      \int\limits_{z_2}^{+\infty} dz_i
                      |\phi_i(\mathbf{B},z_i)|^2 \right)  + \mbox{c.c.}~.   \label{InterfRescTerm}
\end{eqnarray}

\subsection{Elastic antiproton-nucleon scattering amplitude}
For the $\bar p N$ elastic amplitude we neglect the spin- 
and isospin-dependence and apply the following form: 
\begin{equation}
    M_{\bar p N}(\mathbf{q}_t)= 2 i p_{\rm lab} m \sigma_{\bar p p}^{\rm tot}
     (1-i\rho_{\bar p p}) \mbox{e}^{-B_{\bar p p}\mathbf{q}_t^2/2}~,    \label{M_pbarN}
\end{equation}
with $\rho_{\bar p p}=\mbox{Re}M_{\bar p N}(0)/\mbox{Im}M_{\bar p N}(0)$.
The empirical data \cite{PDG2012} tell us that ratio $\rho_{\bar p p}$ quickly
changes sign at $\sqrt{s} \simeq 3-4$ GeV, i.e. just in the region of the $\chi_c$ formation.   
On the other hand, the recent calculations within the Reggeized Pomeron exchange model
\cite{Fiore:2009dz} which seems to agree with empirical data at higher energies 
\cite{Amos:1985wx} predict a smooth behaviour of $\rho_{\bar p p} \simeq -0.05$ 
in the interval $\sqrt{s} \simeq 3-5$ GeV. 
For the slope parameter we choose the value
$B_{\bar p p}=12.5 \pm 1$ GeV$^{-2}$ which is in a good agreement with
empirical slopes at $\sqrt{s} \simeq 3.4-7.0$ GeV 
(or at $p_{\rm lab} \simeq 5-25$ GeV/c) \cite{Antipov:1973tq}.
The total $\bar p p$ cross section has being suitably parametrized
by PDG in \cite{PhysRevD.50.1173}:
\begin{equation}
   \sigma_{\bar p p}^{\rm tot}(p_{\rm lab}) =
   38.4+77.6p_{\rm lab}^{-0.64}+0.26\ln^2(p_{\rm lab})-1.2\ln(p_{\rm lab})~,  \label{sig_pbarp^tot}
\end{equation}
with the beam momentum $p_{\rm lab}$ in GeV/c and the cross section in mb.

\subsection{Formation amplitude $\bar p p \to \chi_J$}

\begin{figure}
\begin{center}
\includegraphics[scale = 0.7]{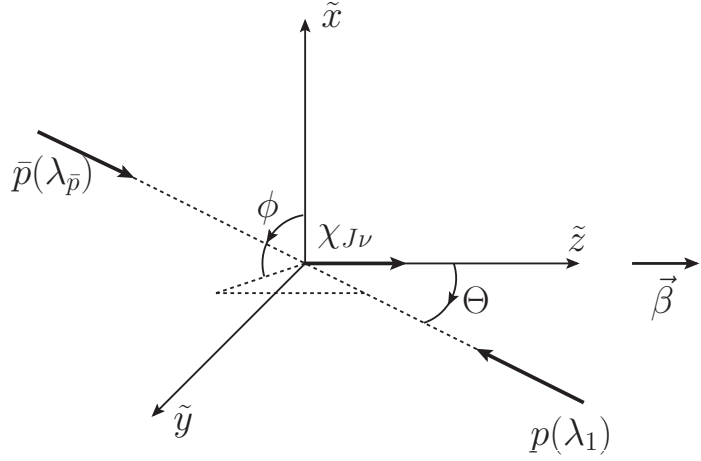}
\end{center}
\caption{\label{fig:elem} Illustration of elementary transition
$\bar p(\lambda_{\bar p}) p(\lambda_1) \to \chi_{J\nu}$, where $\lambda_{\bar p}$,
$\lambda_1$ and $\nu$ are particle helicities.
The picture refers to the c.m. frame of colliding antiproton and proton.
The $\tilde z$-axis is directed along the c.m. velocity $\bvec{\beta}$.}
\end{figure}
The elementary process  $\bar p(\lambda_{\bar p}) p(\lambda_1) \to \chi_{J\nu}$
is depicted in Fig.~\ref{fig:elem} in the $\bar p p$ center-of-mass (c.m.)
frame. Generally, due to a finite transverse momentum of the proton,
the direction of the c.m. velocity 
\begin{equation}
   \bvec{\beta}=\frac{\mathbf{p}_{\bar p}+\mathbf{p}_1}{E_{\bar p}+E_1}  \label{beta}
\end{equation}
does not coincide with the original beam direction. Therefore, 
the transformation from the laboratory frame to 
the coordinate system shown in Fig.~\ref{fig:elem} is obtained 
in the following way. First, we apply the Lorentz boost
from the laboratory frame $(x,y,z)$
to the $\bar p p$ c.m. frame $(x^\prime,y^\prime,z^\prime)$ such that the 
antiproton momentum components become
\begin{equation}
   \mathbf{p}_{\bar p}^\prime = \mathbf{p}_{\bar p}
    - \left( E_{\bar p}
            -\frac{\gamma}{\gamma+1}(\mathbf{p}_{\bar p}\bvec{\beta})
      \right)\gamma\bvec{\beta}~,    \label{p_pbar^prime}
\end{equation}
where $\gamma=1/\sqrt{1-\beta^2}$. Second, we perform a rotation of coordinate 
axes $(x^\prime,y^\prime,z^\prime)$ to the new axes $(\tilde x,\tilde y,\tilde z)$
such that the $\tilde z$-axis becomes alongated with the c.m. velocity $\bvec{\beta}$.
If we denote the polar and azimuthal angles defining the direction of the vector 
$\bvec{\beta}$ in the laboratory frame (or equivalently in the $(x^\prime,y^\prime,z^\prime)$
frame) as $(\Theta_\beta,\phi_\beta)$ then the rotation can be done about the axis
defined by vector $[\mathbf{e}_z^\prime\times\bvec{\beta}]$ by the angle
$\Theta_\beta$, according to the convention of
Refs. \cite{Martin:1984pa,Ridener:1992xr,Jacob:1959at}.
In the resulting coordinate system $(\tilde x,\tilde y,\tilde z)$ the cartesian
components of the antiproton three-momentum are, therefore, given by 
the orthogonal matrix transformation (c.f.\cite{Varshalovich})
\begin{equation}
   \left(
   \begin{array}{l}
      p_{\bar p}^{\tilde x} \\
      p_{\bar p}^{\tilde y} \\
      p_{\bar p}^{\tilde z}
   \end{array}
   \right)
 = \left(
\begin{array}{lll}
   \cos^2\phi_\beta\cos\Theta_\beta+\sin^2\phi_\beta &
   \cos\phi_\beta\sin\phi_\beta(\cos\Theta_\beta-1)  &
    -\cos\phi_\beta\sin\Theta_\beta \\
   \sin\phi_\beta\cos\phi_\beta(\cos\Theta_\beta-1)  &
   \sin^2\phi_\beta\cos\Theta_\beta+\cos^2\phi_\beta &
    -\sin\phi_\beta\sin\Theta_\beta \\
   \sin\Theta_\beta\cos\phi_\beta &
   \sin\Theta_\beta\sin\phi_\beta &
   \cos\Theta_\beta 
\end{array}
 \right) 
\left(
   \begin{array}{l}
      p_{\bar p}^{x^\prime} \\
      p_{\bar p}^{y^\prime} \\
      p_{\bar p}^{z^\prime}
   \end{array}
\right)~.                \label{RotTransform}
\end{equation}

In the notations of Refs. \cite{Martin:1984pa,Ridener:1992xr,Jacob:1959at},
the formation
amplitude of the $\chi_J$-charmonium state with helicity $\nu$ is
\begin{equation}
   \langle J \nu| B |\Theta \phi, \lambda_{\bar p} \lambda_1 \rangle
    =\left(\frac{2J+1}{4\pi}\right)^{1/2} B_{\lambda_{\bar p}\lambda_1}^J 
     D^J_{\nu\lambda}(\phi,\Theta,-\phi) ~,   \label{Bamplitude}
\end{equation}
where $D^J_{\nu\lambda}(\phi,\Theta,-\phi)$ is the rotation matrix,
$\lambda=\lambda_{\bar p}-\lambda_1$ is the net helicity. 
The angles $(\Theta,\phi)$ in Eq. (\ref{Bamplitude})
are the polar and azimuthal angles of the antiproton momentum in
the $(\tilde x, \tilde y, \tilde z)$ coordinate system (see Fig. \ref{fig:elem}),
i.e. $\Theta=\arccos[p_{\bar p}^{\tilde z}
/({p_{\bar p}^{\tilde x}}^2+{p_{\bar p}^{\tilde y}}^2+{p_{\bar p}^{\tilde z}}^2)^{1/2}]$
and $\phi=\arctan(p_{\bar p}^{\tilde y}/p_{\bar p}^{\tilde x})$ ($0 \leq \phi < 2\pi$).
For the zero transverse momentum of the proton the net helicity
is conserved since $D^J_{\nu\lambda}(0,0,0)=\delta_{\nu\lambda}$.
The coefficients $B_{\lambda_{\bar p}\lambda_1}^J$ are normalized as
\begin{equation}
   \sum_{\lambda_{\bar p}\lambda_1} |B_{\lambda_{\bar p}\lambda_1}^J|^2=1~.  \label{NormB}
\end{equation}
The invariant amplitude is proportional to the amplitude (\ref{Bamplitude})
\begin{equation}
    M_{J\nu; \lambda_{\bar p} \lambda_1} 
  = \kappa_J \langle J \nu| B |\Theta \phi, \lambda_{\bar p} \lambda_1 \rangle~, \label{MvsB}
\end{equation}
where the coefficient $\kappa_J$ can be reconstructed from the partial decay 
width $\Gamma_{\chi_J \to \bar p p}$ which gives the relation
\begin{equation}
   \kappa_J = \left(\frac{64 \pi^2 m_J^2 \Gamma_{\chi_J \to \bar p p}}%
{\sqrt{m_J^2-4m^2}}\right)^{1/2}~.      \label{kappa_J}
\end{equation} 
The partial wave amplitudes $B_{\lambda_{\bar p}\lambda_1}^J$ encode the dynamics 
of the charmonium formation. It is, however, possible to obtain some
general relations from the symmetry considerations \cite{Ridener:1992xr}.
It follows from the charge conjugation invariance, that 
$B_{\lambda_{\bar p}\lambda_1}^J=\eta_c(-1)^J B_{\lambda_1\lambda_{\bar p}}^J$,
where $\eta_c=(-1)^{L+S}=1$ is the charge parity of the charmonium 
(for $\chi$-states $L=S=1$). It is convenient to introduce the notations 
$B_0/\sqrt{2} \equiv B_{++}^J$ and $B_1 \equiv B_{+-}^J$ for each $J$.
Then the charge conjugation invariance leads to the condition
$|B_{-+}^J|^2=|B_{+-}^J|^2 = |B_1|^2$.
The parity invariance of the amplitude (\ref{Bamplitude}) gives the relation 
$B_{\lambda_{\bar p}\lambda_1}^J=\eta_p (-1)^J B_{-\lambda_{\bar p},-\lambda_1}^J$,
where $\eta_p=(-1)^{L+1}=1$ is the charmonium parity. This results in 
to relations $|B_{++}^J|^2=|B_{--}^J|^2 = |B_0|^2/2$.
Moreover, in the case of $\chi_1$, the charge conjugation invariance leads to the
condition $B_0=0$. The partial wave amplitudes $B_0$ and $B_1$  are
normalized as
\begin{equation}
   2|B_1|^2 + |B_0|^2 = 1~.                  \label{NormB01}
\end{equation}
The recent experimental data \cite{Ambrogiani:2001jw} for the angular distributions from
the $\bar p p \to \chi_{c2} \to J/\psi \gamma \to e^+ e^- \gamma$
decay provide the value $|B_0|^2=0.13\pm0.08$.
The smallness of the transition amplitude for the net helicity zero
can be understood as a signature of hadronic helicity conservation
for exclusive processes within perturbative QCD with massless 
quarks and spin-1 gluons \cite{Brodsky:1981kj}.

In calculations of the products of the matrix elements for the processes
$\bar p p \to \chi_{J\nu}$ and $\bar p p \to \chi_{J_1\nu}$ we will assume
for simplicity the proton longitudinal momentum to be $p_1^z=(\Delta_0^J+\Delta_0^{J_1})/2$.
This approximation is good enough for the present exploratory
studies.
(More rigorously, in the first amplitude on should set $p_1^z=\Delta_0^J$ 
and in the second amplitude $p_1^z=\Delta_0^{J_1}$.)
Then, the azimuthal angle $\phi$ will cancel in 
the final results for the squares of the matrix elements.
This can be seen if we use the property of the rotation matrix
(c.f. \cite{Varshalovich}) 
\begin{equation}
   D_{MM^\prime}^J(\alpha,\beta,\gamma)=\mbox{e}^{-i\alpha M}
   d_{MM^\prime}^J(\beta)\mbox{e}^{-i\gamma M^\prime}  \label{Dproperty}
\end{equation}
with $d_{MM^\prime}^J(\beta)$ being the real-valued functions.
The consequence is that the formulas (\ref{DirectTerm}),(\ref{InterfTerm}),%
(\ref{DiagRescTerm}),(\ref{NondiagRescTerm}) and (\ref{InterfRescTerm})
derived earlier depend on the combinations
\begin{eqnarray}
   M_{J_1\nu; \lambda_{\bar p} \lambda_1}(\mathbf{q}_t) 
   M_{J\nu; \lambda_{\bar p} \lambda_1}^*(\mathbf{q}_t)   
   &=& \kappa_{J_1} \kappa_J \frac{\sqrt{(2J_1+1)(2J+1)}}{4\pi}
     B_{\lambda_{\bar p}\lambda_1}^{J_1} B_{\lambda_{\bar p}\lambda_1}^{J*}  \nonumber \\
   && \times  d_{\nu\lambda}^{J_1}(\Theta) d_{\nu\lambda}^{J}(\Theta)~.   \label{SimpleProduct}
\end{eqnarray}
The phases of the helicity amplitudes $B_{\lambda_{\bar p}\lambda_1}^{J}$ are unknown.
We will fix $B_0=1$ for $J=0$ and $B_1=1/\sqrt{2}$ for $J=1$.
In most calculations we will assume the zero phases of the $J=2$ helicity amplitudes, i.e.
$B_0=\sqrt{0.13}$ and $B_1=\sqrt{0.87/2}$.
However, we will also test several different choices of the phases of $B_0$ and $B_1$ for $J=2$
This will influence the interference terms (\ref{InterfTerm}),(\ref{InterfRescTerm}) only.

\subsection{Transition amplitudes $\chi_{J_1} N \to \chi_J N$}

Following \cite{Gerland:1998bz} we decompose the internal $c\bar c$
wave function of the physical $\chi_{J\nu}$-charmonium state
in the basis of wave functions with fixed orbital ($L_z$)
and spin ($S_z$) magnetic quantum numbers as
\begin{equation}
    |J\nu\rangle=\sum_{L_z,S_z} |1L_z;1S_z\rangle \langle 1L_z;1S_z|J\nu\rangle~,
                        \label{Jnu_decomp}
\end{equation}
where $z$-axis is directed along the charmonium momentum
in the target nucleus rest frame (Fermi motion is neglected here). 
$\langle 1L_z;1S_z|J\nu\rangle$ are the Clebsch-Gordan coefficients.
To avoid misunderstanding, we speak here about the {\it internal}
orbital angular momentum of a $c\bar c$-pair. (The projection
of the c.m. orbital momentum of the $c\bar c$-pair on the $z$-axis
is identically zero.) Assuming that the interaction does not change
the internal spin and angular momentum of the $c\bar c$-pair we
can approximate the rescattering amplitude as
\begin{eqnarray}
    \langle J\nu |\hat S|J_1\nu\rangle=\sum_{L_z,S_z} \langle J\nu |1L_z;1S_z\rangle 
    \langle 1L_z;1S_z|\hat S|1L_z;1S_z\rangle  \langle 1L_z;1S_z|J_1\nu\rangle~,
                        \label{Jnu_S}
\end{eqnarray} 
where the symbols of the initial and final nucleons are dropped for brevity.
Assuming that the ratios between diagonal and nondiagonal transitions
do not change with increasing transverse momentum transfer
Eq.(\ref{Jnu_S}) can be rewritten for the invariant matrix elements:
\begin{eqnarray}
    M_{J\nu;J_1\nu}(\mathbf{q}_t)= \mbox{e}^{-B_{\chi N}\mathbf{q}_t^2/2}
     \sum_{L_z,S_z} \langle J\nu |1L_z;1S_z\rangle 
      M_{L_zS_z}(0) \langle 1L_z;1S_z|J_1\nu\rangle~.   \label{Jnu_M}
\end{eqnarray}
In the two-gluon exchange mechanism $B_{\chi N}\simeq3$ GeV$^{-2}$ 
for the discussed energy range \cite{Gerland:2005ca}.
For the forward scattering amplitudes at fixed $L_z$ and $S_z$ we have
\begin{equation}
    M_{L_zS_z}(0) = 2 i p_{\rm lab} m \sigma_{L_zS_z}^{\rm tot}
                    (1-i\rho_{\chi N})~.                \label{M_LzSz(0)}
\end{equation}
Here $\rho_{\chi N}=\mbox{Re}M_{L_zS_z}(0)/\mbox{Im}M_{L_zS_z}(0)$.
From the soft Pomeron exchange one has $\rho_{\chi N} \simeq 0.15$, while
pQCD gives $\rho_{\chi N} \simeq 0.3$.  In numerical calculations
we have chosen $\rho_{\chi N}=0.22$, i.e. the average of these two values,
since the sensitivity to $\rho_{\chi N}$ in the interval 0.15-0.3 turns out 
to be quite modest (see the right panel of Fig.~\ref{fig:ratio_plabDep_sum}
below).
 
The most important input of our calculations are the cross sections 
$\sigma_{L_zS_Z}^{\rm tot} \equiv \sigma_{L_z}$ which have been calculated 
in Ref. \cite{Gerland:1998bz} on the basis of the nonrelativistic quark model
and the QCD factorization theorem. The following values have been obtained
in \cite{Gerland:1998bz}: $\sigma_0=6.8$ mb and $\sigma_{\pm1}=  15.9$ mb.
The cross sections differ by approximately a factor of two, since the transverse
size squared of the $c\bar c$ configuration with $L_z=\pm1$ is two times larger as
compared to the one of the configuration with $L_z=0$. The exact ratio $\sigma_1/\sigma_0$
deviates from two, because the cross sections are obtained in Ref. \cite{Gerland:1998bz}
by weighting the probability density distribution of the relative quark coordinate
with the transverse-size-dependent interaction cross section of a $c\bar c$ pair
with a nucleon. The latter cross section was evaluated in \cite{Gerland:1998bz}
based on nonperturbative QCD.

In Table~\ref{tab:Amplitude} we list the transition amplitudes $M_{J\nu;J_1\nu}(0)$
for the different values of $J$, $\nu$ and $J_1$. 
\begin{table}[htb]
\caption{\label{tab:Amplitude} Transition amplitude for different initial ($J_1$) 
and  final ($J$) total angular momenta and helicities ($\nu$)
of the $\chi_{c}$. For $\nu < 0$ one should use the relation 
$M_{J-\nu;J_1-\nu}(0)=(-1)^{J+J_1}M_{J\nu;J_1\nu}(0)$. The quantities
$M_{L_z} \equiv M_{L_zS_z}(0),~L_z=0,1$ denote the amplitudes with
fixed value of the $z$-component of the orbital angular momentum neglecting
their spin dependence.}
\begin{center}
\begin{tabular}{cccc}
\hline
$J$ & $\nu$ & $J_1$ & $M_{J\nu;J_1\nu}(0)$ \\ 
\hline
0   & 0     & 0   & $(2M_1 + M_0)/3$ \\
0   & 0     & 1          & 0 \\
0   & 0     & 2   & $\sqrt{2}(M_1-M_0)/3$ \\
1   & 0     & 1          & $M_1$ \\
1   & 0     & 2          & 0 \\
1   & 1     & 1   & $(M_1 + M_0)/2$ \\
1   & 1     & 2   & $(M_1 - M_0)/2$ \\
2   & 0     & 2   & $(M_1 + 2M_0)/3$ \\
2   & 1     & 2   & $(M_1 + M_0)/2$ \\
2   & 2     & 2          & $M_1$ \\
\hline
\end{tabular}
\end{center}
\end{table}
The nondiagonal transition amplitudes between physical $\chi_c$ states
are proportional to the difference between the amplitudes with $L_z=1$ 
and $L_z=0$. Hence, the nondiagonal transitions are governed by the
difference $\sigma_1-\sigma_0$ which turns out to be nonzero according
to the quark model predictions on the structure of the $\chi_c$ states and
QCD factorization theorem.  

\subsection{Occupation numbers}
\label{OccNum}

The squares of $\chi_J$ production amplitudes on a nucleus 
(c.f. Eq. (\ref{IAterm}) etc.) are proportional to the coordinate-
and momentum-dependent occupation number $f_1(\mathbf{X},\mathbf{p})$
of the struck proton which is formally defined as the Wigner function
(\ref{WF}). The cross section on the nucleus (\ref{dsigma}) includes 
the sum over all possible struck protons ($N_1$). Thus, 
the cross section depends on the total proton occupation number
$n(\mathbf{X},\mathbf{p})$ defined as
\begin{equation}
    2 n(\mathbf{X},\mathbf{p}) = \sum_{N_1} f_1(\mathbf{X},\mathbf{p})~. \label{TotalOccNumber}
\end{equation}
By introducing the factor of 2 we assumed the spin saturation of the proton system
in the nucleus. In the present work we will use a simple expression for
$n(\mathbf{X},\mathbf{p})$, which is based on the local Fermi distribution 
but takes into account the corrections due to the short range NN correlations 
(SRC):
\begin{equation}
   n(\mathbf{X},\mathbf{p}) = (1-P_2) \Theta(p_F-p) 
            + \frac{(2\pi)^3}{2} \rho_p a_2 |\psi_D(p)|^2 \Theta(p-p_F)~. \label{nWithTail}
\end{equation}
Here $\rho_p(\mathbf{X})$ is the proton density; $p_F(\mathbf{X})=(3\pi^2\rho_p)^{1/3}$ is the 
proton Fermi momentum; and $P_2 \simeq 0.25$ is the proton fraction above Fermi surface
\cite{Frankfurt:1981mk,Frankfurt:2008zv}.
The deuteron wave function $\psi_D(p)$ in the momentum representation
is normalized as 
\begin{equation}
   4\pi \int\limits_0^{+\infty} dp p^2 |\psi_D(p)|^2 = 1~.   \label{psi_D_Norm}
\end{equation}
The coefficient $a_2(\mathbf{X})$ is chosen from from the condition
\begin{equation}
    P_2 = 4 \pi a_2 \int\limits_{p_F}^{+\infty} dp p^2 |\psi_D(p)|^2~.  \label{P_2}
\end{equation}
For the deuteron wave function we take the result of calculations with the Paris
potential \cite{Lacombe:1980dr}.

Overall, the in-medium effects should grow with the mass number of a target nucleus.
Hence we selected the $^{208}$Pb nucleus for the numerical studies below.
The density distributions of protons and neutrons have been taken in the two-parameter
Fermi parameterization as described in \cite{Larionov:2013axa}.    

\section{Numerical results}
\label{results}

We calculate the transverse momentum differential cross sections of
the $\chi_{J\nu}$ charmonium production
\begin{equation}
   \frac{d\sigma_{\bar p A \to \chi_{J\nu} (A-1)^*}}{d^2k_t} =
    \frac{|M|^2}{16\pi^2p_{\rm lab}^2}~,    \label{dsigdkt}
\end{equation}
which can be obtained by integrating Eq.(\ref{dsigma}) over the longitudinal
momentum $k^z$ and replacing $\sqrt{(E_{\bar p}+E_1)^2-k_t^2-m_J^2}$ by $p_{\rm lab}$
at the final step.
$|M|^2$ stands for the full matrix element squared for the charmonium production 
on the nucleus.
It is important to note, that in deriving Eq.(\ref{dsigdkt}) we implicitly assumed
that the contribution of negative $k^z$ is strongly suppressed by the rescattering 
matrix elements which enter in $|M|^2$. This allowed us to limit the integration 
to the positive values of $k^z$ only. The cross section (\ref{dsigdkt})
is invariant with respect to the change $\nu \to -\nu$, as can be seen from explicit
expressions for the different contributions to $|M|^2$ in the previous section.

\begin{figure}
\includegraphics[scale = 0.6]{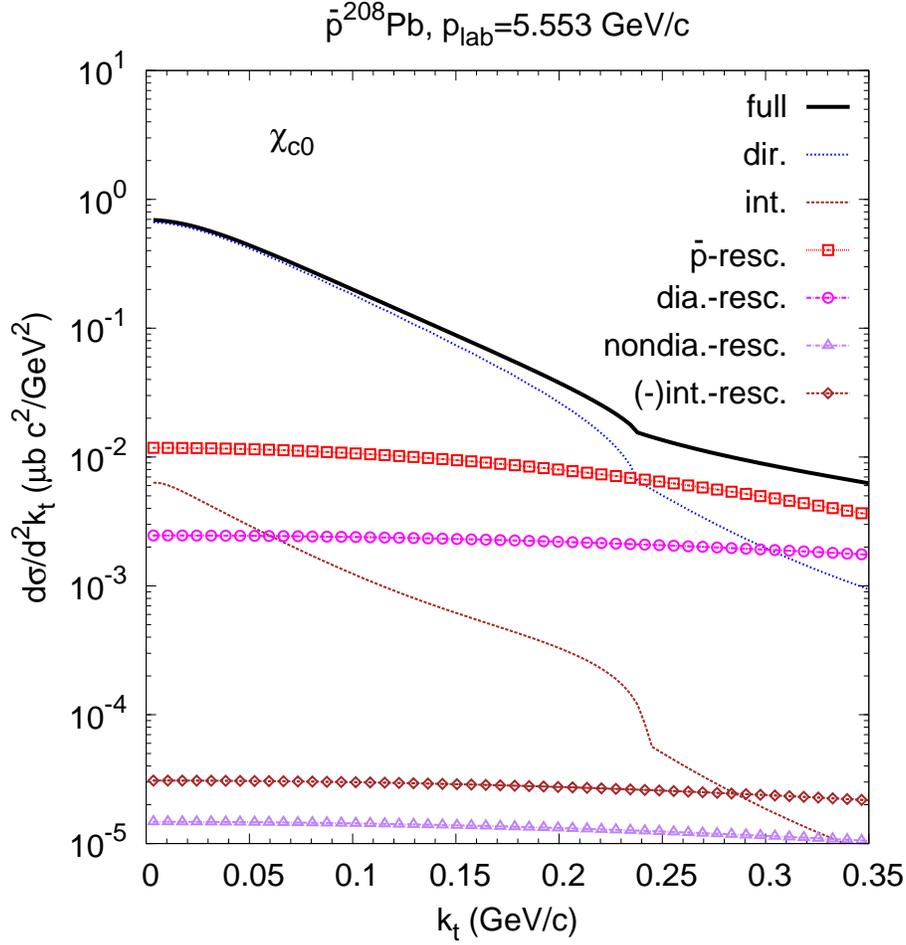}
\caption{\label{fig:d2sig_00}(Color online) Transverse momentum dependence of
the differential $\chi_{c0}$ production cross section (\ref{dsigdkt})
in $5.553$ GeV/c beam momentum  antiproton interactions with the nucleus 
$^{208}$Pb. Full calculation including all contributions
to the matrix element is shown by the solid line. Other lines show
the partial contributions of the different terms.
Direct term (\ref{DirectTerm}) -- blue dotted line.
Interference term (\ref{InterfTerm}) -- brown dashed line.
Antiproton rescattering term (\ref{PbarRescTerm}) -- red squares.
Diagonal rescattering term (\ref{DiagRescTerm}) -- magenta circles.
Nondiagonal rescattering term (\ref{NondiagRescTerm}) -- purple triangles.
Interference rescattering term (\ref{InterfRescTerm}) 
-- brown diamonds (contributes with "-" sign).}
\end{figure}

\begin{figure}
\includegraphics[scale = 0.6]{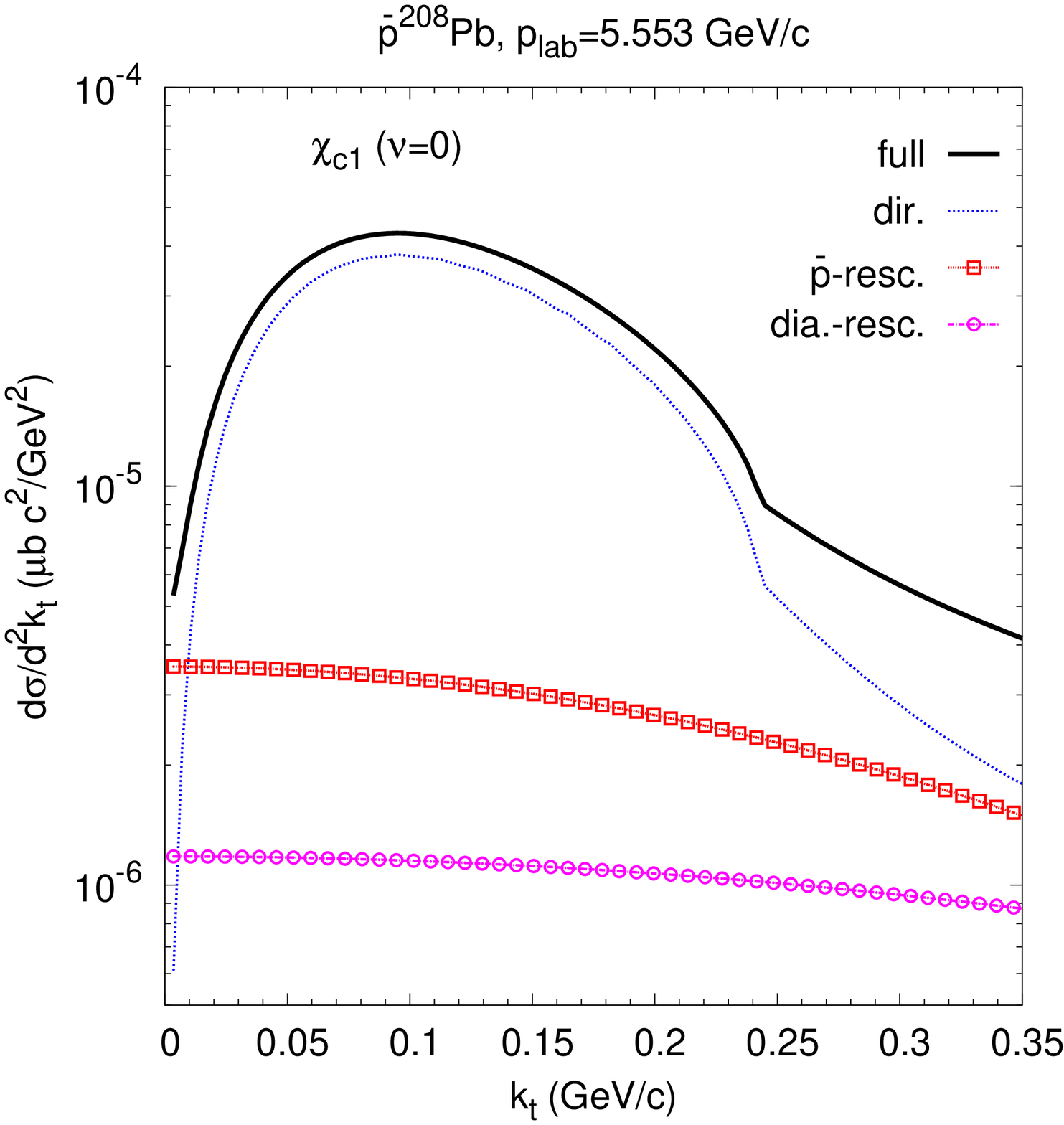}
\caption{\label{fig:d2sig_10}(Color online) Same as Fig.~\ref{fig:d2sig_00},
but for $\chi_{c1}$ production with helicity $\nu=0$.}
\end{figure}

\begin{figure}
\includegraphics[scale = 0.6]{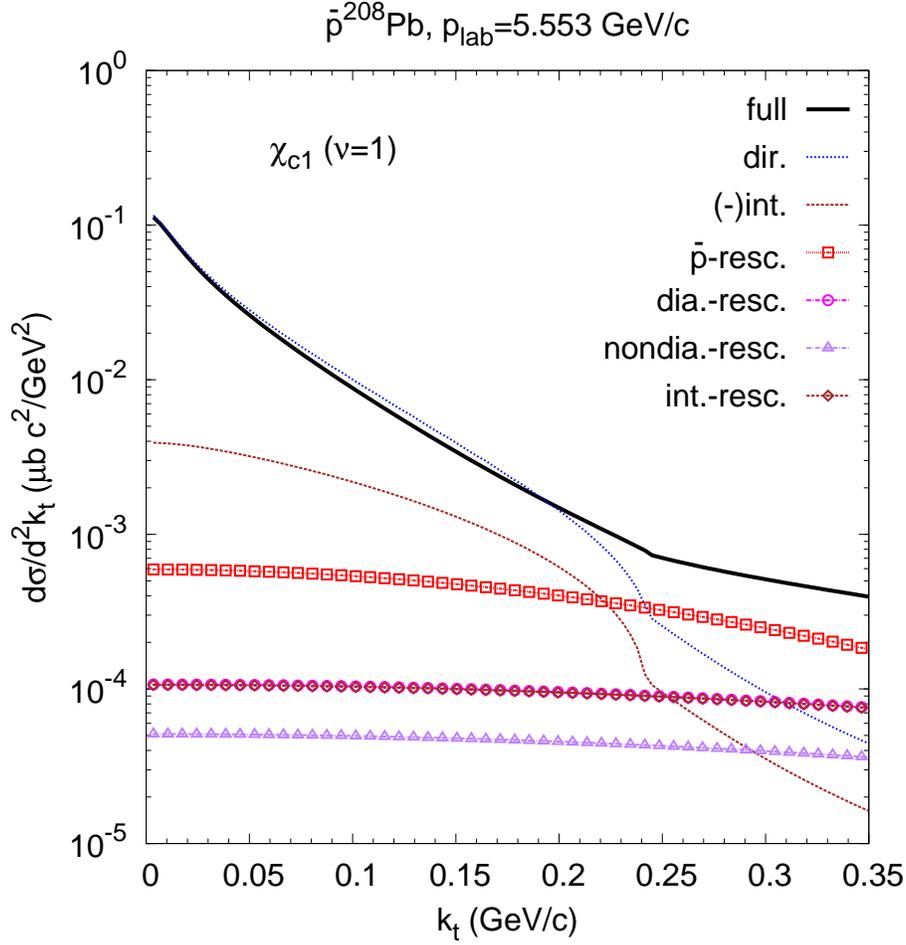}
\caption{\label{fig:d2sig_11}(Color online) Same as Fig.~\ref{fig:d2sig_00},
but for $\chi_{c1}$ production with helicity $\nu=1$. The interference term
(\ref{InterfTerm}) contributes with "-" sign. The contributions 
of the diagonal and interference rescattering terms (\ref{DiagRescTerm}),
(\ref{InterfRescTerm}) almost coincide with each other.}
\end{figure}

\begin{figure}
\includegraphics[scale = 0.6]{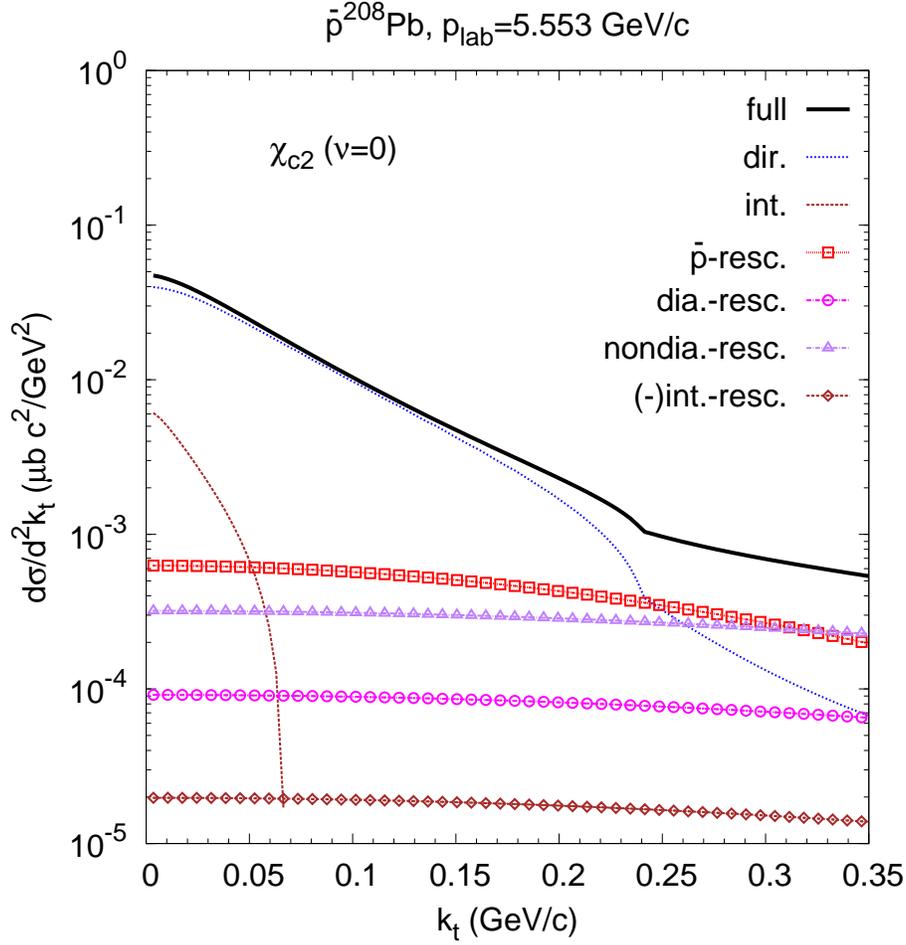}
\caption{\label{fig:d2sig_20}(Color online) Same as Fig.~\ref{fig:d2sig_00},
but for $\chi_{c2}$ production with helicity $\nu=0$. The interference 
rescattering term (\ref{InterfRescTerm}) contributes with "-" sign.}
\end{figure}

\begin{figure}
\includegraphics[scale = 0.6]{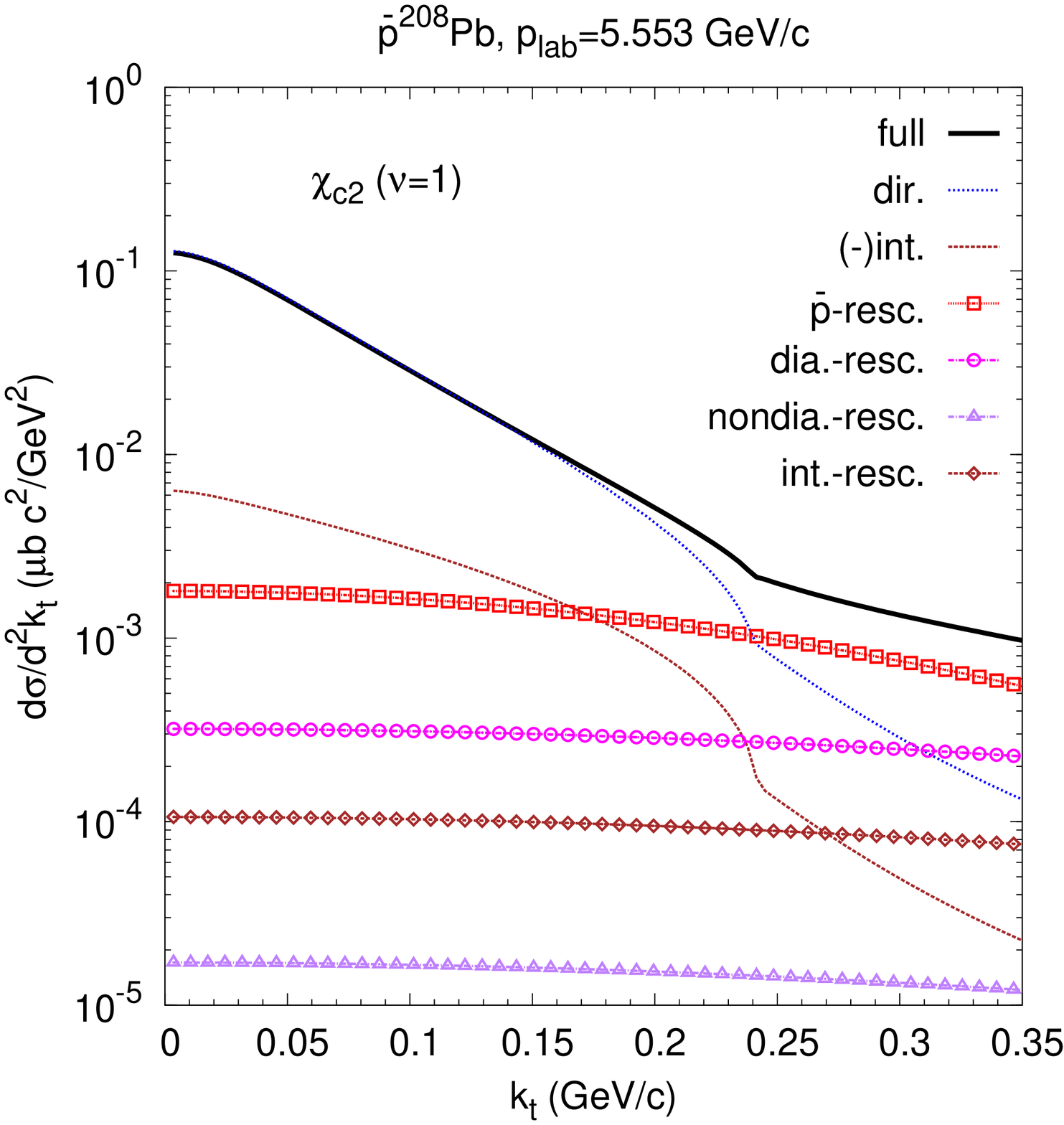}
\caption{\label{fig:d2sig_21}(Color online) Same as Fig.~\ref{fig:d2sig_00},
but for $\chi_{c2}$ production with helicity $\nu=1$. The interference term 
(\ref{InterfTerm}) contributes with "-" sign.}
\end{figure}

\begin{figure}
\includegraphics[scale = 0.6]{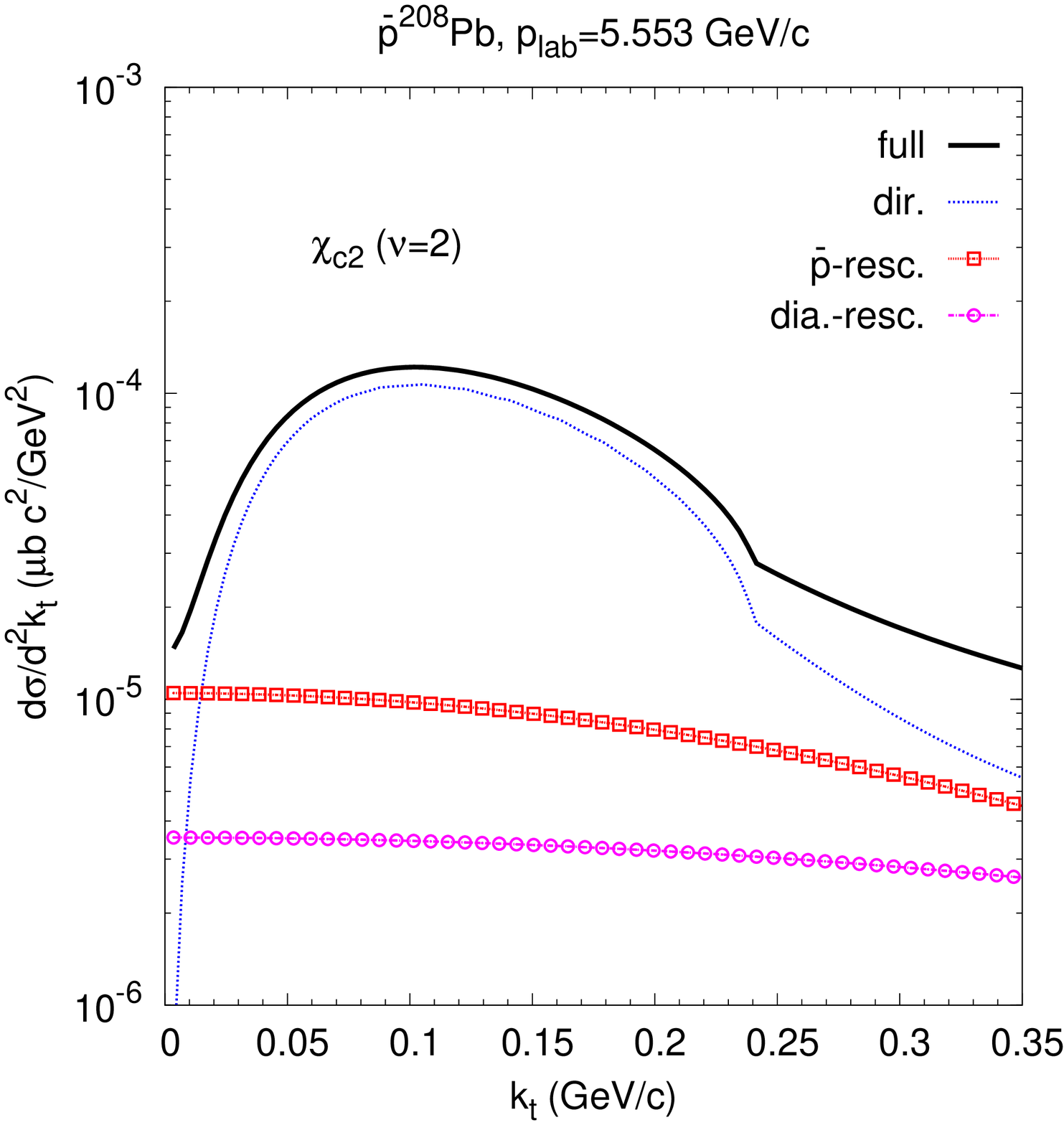}
\caption{\label{fig:d2sig_22}(Color online) Same as Fig.~\ref{fig:d2sig_00},
but for $\chi_{c2}$ production with helicity $\nu=2$.}
\end{figure}

Figures \ref{fig:d2sig_00}-\ref{fig:d2sig_22} show the transverse momentum differential
$\chi_c$-charmonia production cross sections with the different total angular momenta
$J$ and helicities $\nu$. The calculations were performed at an antiproton beam momentum
of $p_{\rm lab}=5.553$ GeV/c corresponding to on-shell $\chi_{c1}$ formation in
$\bar p p$ collisions.
The kinks in the $k_t$-dependence at $k_t \simeq 0.25$ GeV/c are caused by the sharp change
in the momentum dependence of the occupation numbers at the Fermi momentum as discussed 
in the previous section.

For the states $\chi_{00}$, $\chi_{11}$, $\chi_{20}$ and $\chi_{21}$,
whose formation is allowed in $\bar p p$ collisions, 
the cross sections at low $k_t$ are dominated by the direct term (\ref{DirectTerm})
and at $k_t > 0.25$ GeV/c -- by the term with antiproton rescattering (\ref{PbarRescTerm}).
The latter makes the large excess above the SRC tail of the direct term.

The "exotic" states $\chi_{10}$ and $\chi_{22}$ can not be formed in $\bar p p$ 
collisions and are, therefore, strongly suppressed in antiproton-nucleus collisions.
Their production at small $k_t$ is mainly caused by the antiproton rescattering term
and at $k_t > 0.25$ GeV/c -- by the SRC tail of the direct term.
In the latter case the transverse momentum is provided by the target proton.
Hence the charmonium spin quantization axis does not coincide with the beam direction 
anymore ($\Theta > 0$ in Eq.(\ref{SimpleProduct})). As the consequence, the charmonium 
helicity may deviate from the difference of the antiproton and proton helicities.
The production cross sections of the "exotic" $\chi_{10}$ and $\chi_{22}$ states 
on the nucleus are, however, several orders of magnitude lower than for the other
"nonexotic" states $\chi_{00}$, $\chi_{11}$, $\chi_{20}$ and  $\chi_{21}$.

We will discuss now the nondiagonal transitions. Note, first, that such transitions
do not contribute to the $\chi_{10}$ and $\chi_{22}$ production as one can see from
Table~\ref{tab:Amplitude}. On the other hand, the nondiagonal transitions
$11 \leftrightarrow 21$ and $00 \leftrightarrow 20$ do contribute the production
of the respective $\chi_{J\nu}$ states. In particular, the transition $00 \to 20$
influences the $\chi_{20}$ production at low transverse momenta significantly. 
This is caused by the large partial partial width 
$\Gamma_{\chi_{c0} \to \bar p p} \simeq 2.3$ keV as compared to
$\Gamma_{\chi_{c1} \to \bar p p} \simeq 0.06$ keV and
$\Gamma_{\chi_{c2} \to \bar p p} \simeq 0.14$ keV.
As a result, the cross section of $\chi_{20}$ production is enhanced by $\sim 20\%$ at
small $k_t$ due to the interference term (\ref{InterfTerm}).

\begin{figure}
\includegraphics[scale = 0.6]{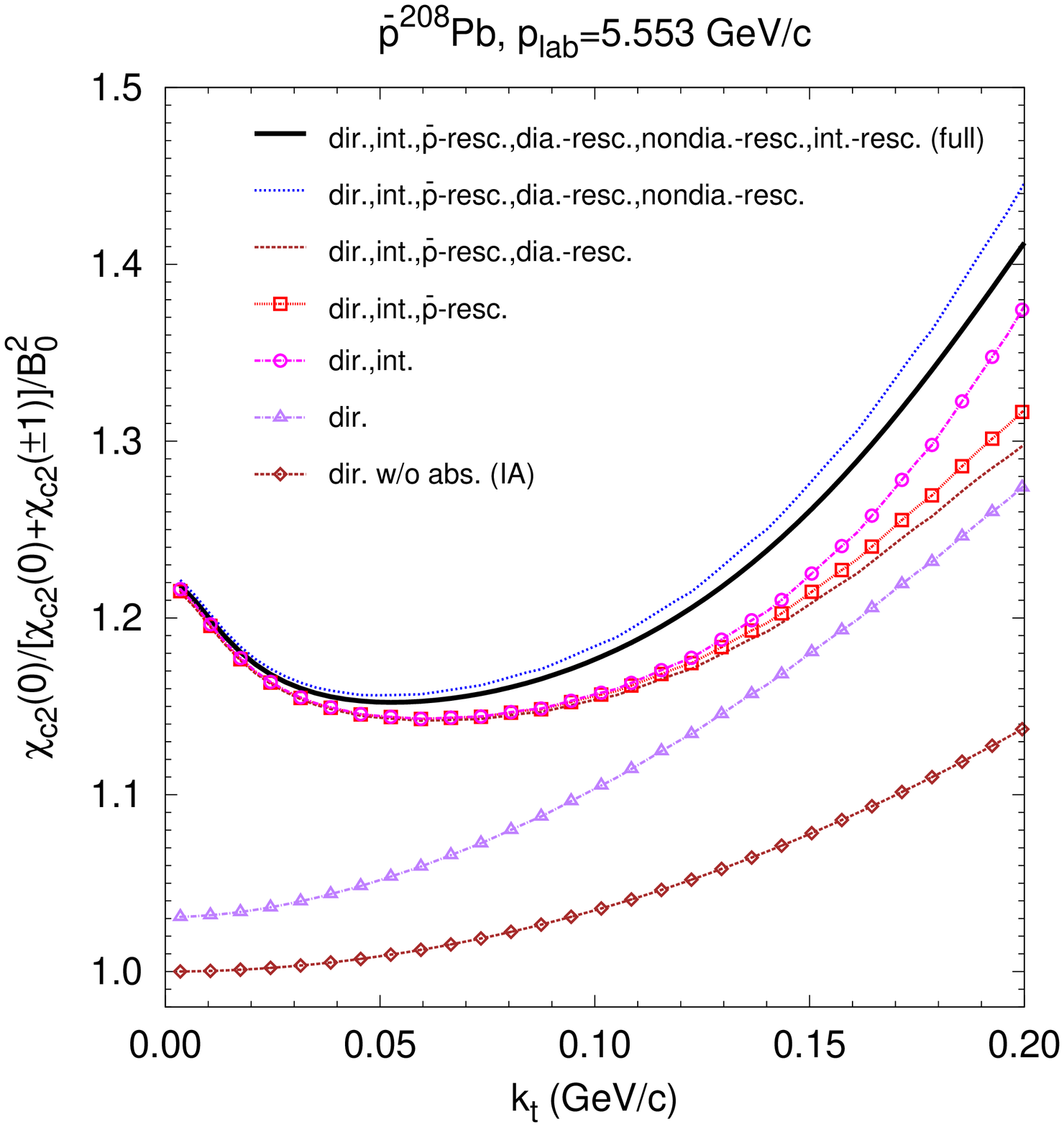}
\caption{\label{fig:ratio_5.553gevc}(Color online) The relative contribution ${\cal R}$
(see Eq.(\ref{calR})) of the $\chi_{c2}$ production with helicity $\nu=0$ to the total 
$\chi_{c2}$ production in antiproton collisions at 5.553 GeV/c with $^{208}$Pb nucleus
vs transverse momentum. The normalization is performed on the same contribution 
$|B_0|^2=0.13$ in the nonpolarized $\bar p p$ collisions.}
\end{figure}

This is better seen in Fig.~\ref{fig:ratio_5.553gevc} which shows the normalized ratio
\begin{equation}
   {\cal R} = \frac{\chi_{20}}{(\chi_{20}+2\chi_{21})|B_0|^2}   \label{calR}
\end{equation}
as a function of $k_t$.
In the abscence of any in-medium effects (impulse approximation, Eq.(\ref{IAterm}))
we have ${\cal R}=1$ at $k_t=0$. Including absorption (direct term, Eq.(\ref{DirectTerm}))
increases ${\cal R}$ by about $5\%$, which reflects the genuine color filtering
effect. Indeed, one can see from Table~\ref{tab:Amplitude} that the absorption
cross section of the $\chi_{21}$ state is slightly larger than the absorption
cross section of the $\chi_{20}$ state (since $\mbox{Im} M_1 > \mbox{Im} M_0$).
The interference term (\ref{InterfTerm}) leads to an additional and quite 
significant enhancement of ${\cal R}$, so that it reaches $\sim 20\%$.
The enhancement is not affected by the rescattering terms, which do not 
influence the ratio ${\cal R}$ at small $k_t$, practically.

\begin{figure}
\includegraphics[scale = 0.6]{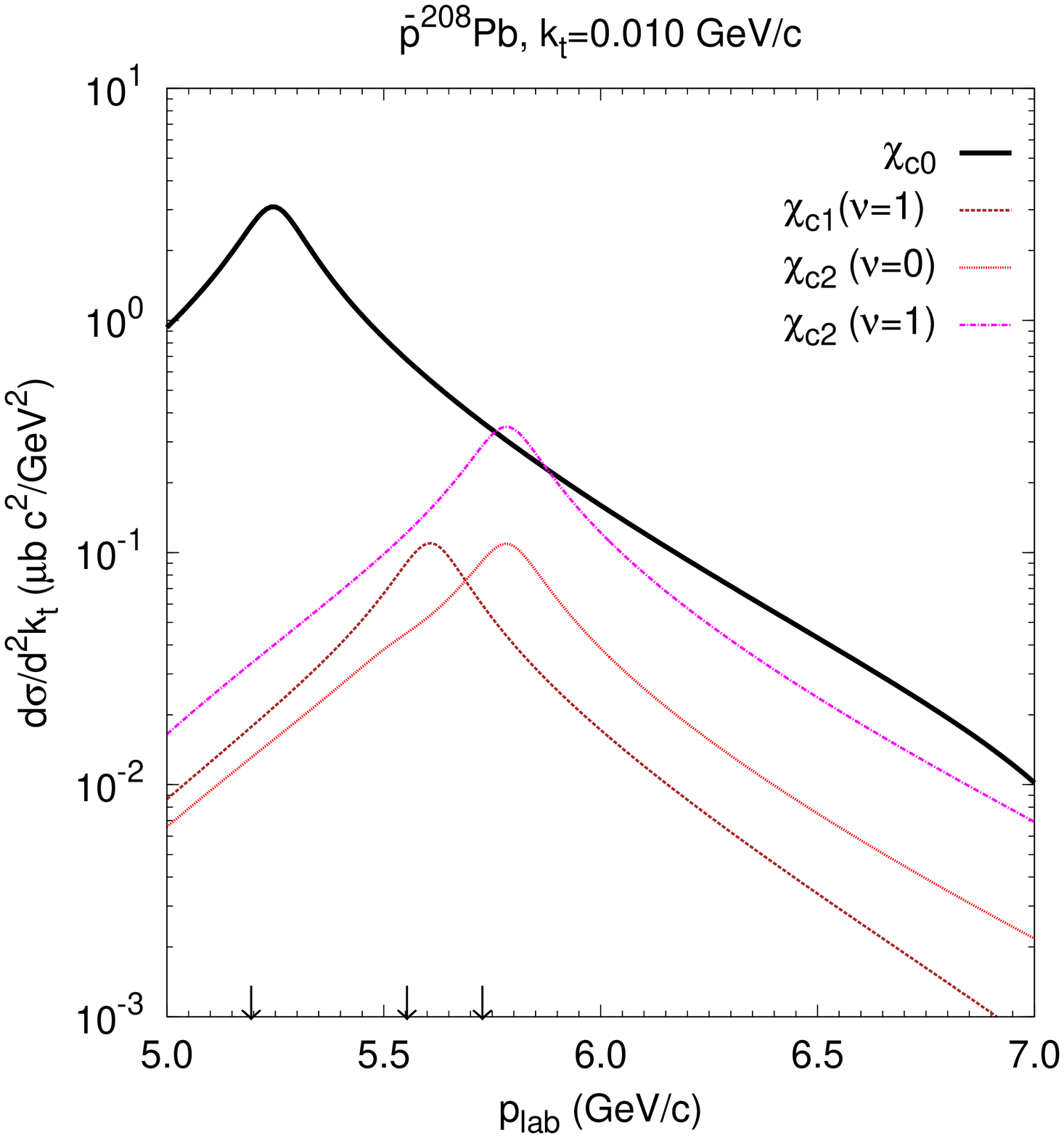}
\caption{\label{fig:d2sigd2kt_vs_plab}(Color online) Transverse momentum
differential cross section of $\chi_{c0}$, $\chi_{c1}$ and $\chi_{c2}$
production with different helicities plotted vs beam 
momentum  at the fixed $k_t=0.010$ GeV/c.
For orientation, vertical arrows show the beam momenta 
of the on-shell $\chi_{c0}$, $\chi_{c1}$ and $\chi_{c2}$ formation
in $\bar p p$ collisions ($p_{\rm lab}=5.194,~5.553$ and 5.727 GeV/c,
respectively). Note, that the cross sections are peaked at slightly higher
beam momenta due the finite value of the nucleon binding energy
($7.9$ MeV for the $^{208}$Pb nucleus).}
\end{figure}
In Fig.~\ref{fig:d2sigd2kt_vs_plab} we display the beam momentum dependence of the transverse
momentum differential cross sections for the "nonexotic" $\chi_c$-states 
at low transverse momentum 
\footnote{
We have chosen a small but finite value of
$k_t$ in order to avoid the singularities in the space integral of the occupation
numbers for the on-shell charmonia production at $k_t=0$.
The singularities appear from the volume integration in the direct term (\ref{DirectTerm})
when $k_t=0$ and $\Delta^0_J=0$ and in the interference term (\ref{InterfTerm})
when $k_t=0$ and $\Delta^0_J+\Delta^0_{J_1}=0$. This is
because of the two-parameter Fermi distribution density tail which is infinite
in radius. It follows from Eqs. (\ref{TotalOccNumber}),(\ref{nWithTail})
that $f_1(\mathbf{X},0)=(2/Z) n(\mathbf{X},0)=(2/Z)(1-P_2)=\mbox{const}$
independent of position $\mathbf{X}$. The singularities are integrable, i.e.
the integration of the cross section (\ref{dsigdkt}) over $d^2k_t$ gives 
the finite result. In the rescattering terms 
(\ref{PbarRescTerm})-(\ref{InterfRescTerm}) singularities do not
appear due to the integration over $d^2t_2$.}.
Due to Fermi motion and SRC-correlations there is a strong overlap of the 
$\chi_{c0}$, $\chi_{c1}$ and $\chi_{c2}$ production in the considered region
of beam momenta. This makes possible the interference between these states,
since the phase multiplication factor $\Delta_{J_1}^0-\Delta_{J}^0$ in 
Eq. (\ref{InterfTerm}) is small.   

\begin{figure}
\includegraphics[scale = 0.4]{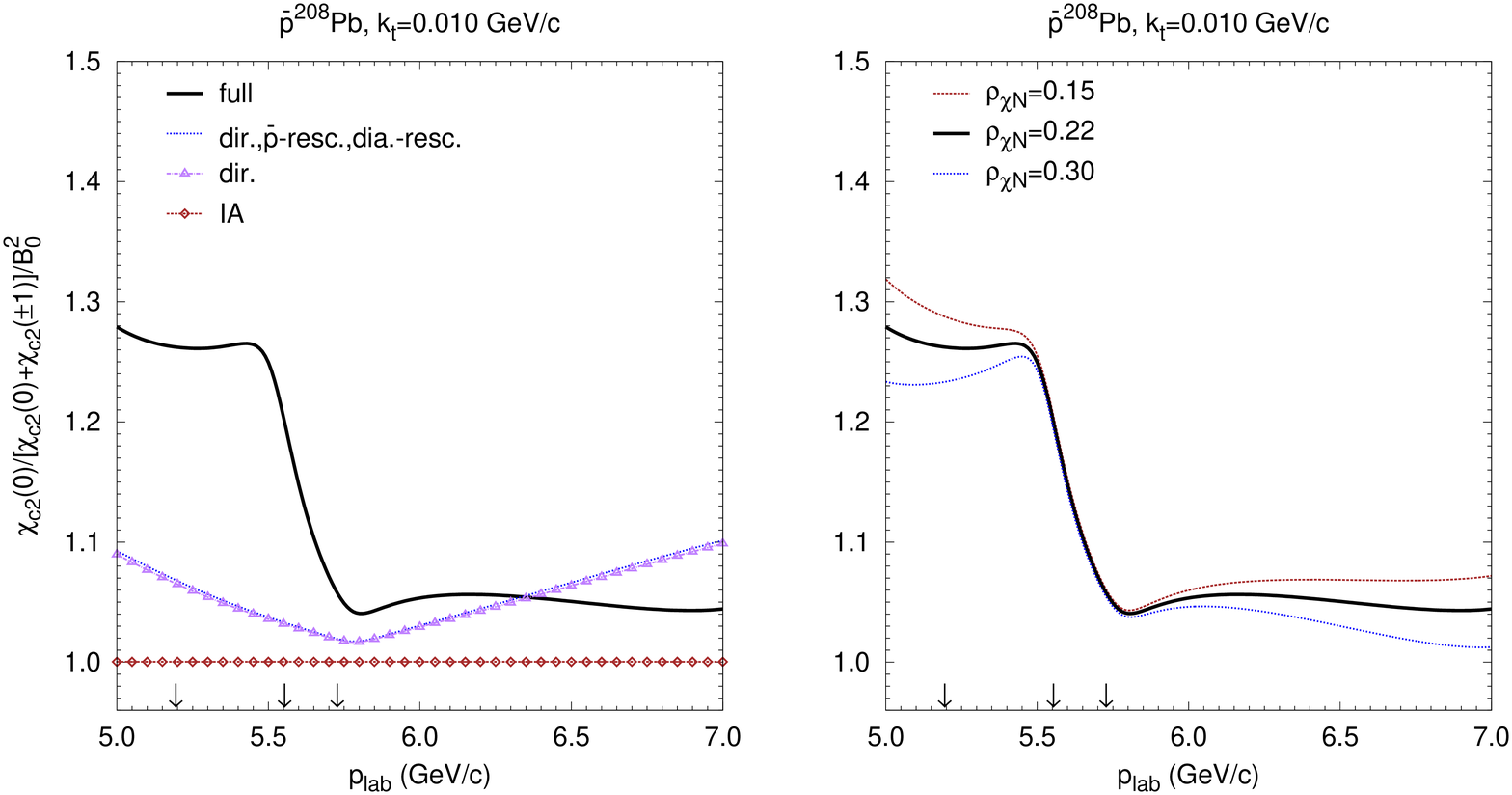}
\caption{\label{fig:ratio_plabDep_sum}(Color online) The normalized fraction of
the $\chi_{c2}$ production with helicity $\nu=0$ (see Eq.(\ref{calR})) at 
$k_t=0.010$ GeV/c on the $^{208}$Pb nucleus as a function of the antiproton 
beam momentum.
Vertical arrows show the beam momenta of the on-shell 
$\chi_{c0}$, $\chi_{c1}$ and $\chi_{c2}$ formation
in $\bar p p$ collisions ($p_{\rm lab}=5.194,~5.553$ and 5.727 GeV/c,
respectively).
Left panel shows the calculations with fixed value of $\rho_{\chi N}=0.22$
including combinations of the different terms as indicated.
Right panel shows the sensitivity of the full calculation
to the choice of parameter $\rho_{\chi N}$ (see Eq.(\ref{M_LzSz(0)}))}.
\end{figure}

Figure~\ref{fig:ratio_plabDep_sum} shows the beam momentum dependence 
of the ratio ${\cal R}$ at $k_t=0.010$ GeV/c.
The ratio reaches flat maximum at the beam momentum of about 5.5 GeV/c
corresponding to $\Delta_0^0+\Delta_2^0=0$, where the occupation number
in the interference term (\ref{InterfTerm}) is maximal. We also see 
that ${\cal R}$ drops quickly with increasing beam momentum between    
$\simeq 5.50$ GeV/c and $\simeq5.56$ GeV/c. Our results reveal a modest
sensitivity to the choice of the ratio of the real and imaginary parts
of the $\chi N$-scattering amplitude 
(right panel of Fig.~\ref{fig:ratio_plabDep_sum}). However, this sensitivity
reaches at most $\sim10\%$ and is visible only for far-off-shell 
$\chi_{c2}$-production.

\begin{figure}
\includegraphics[scale = 0.6]{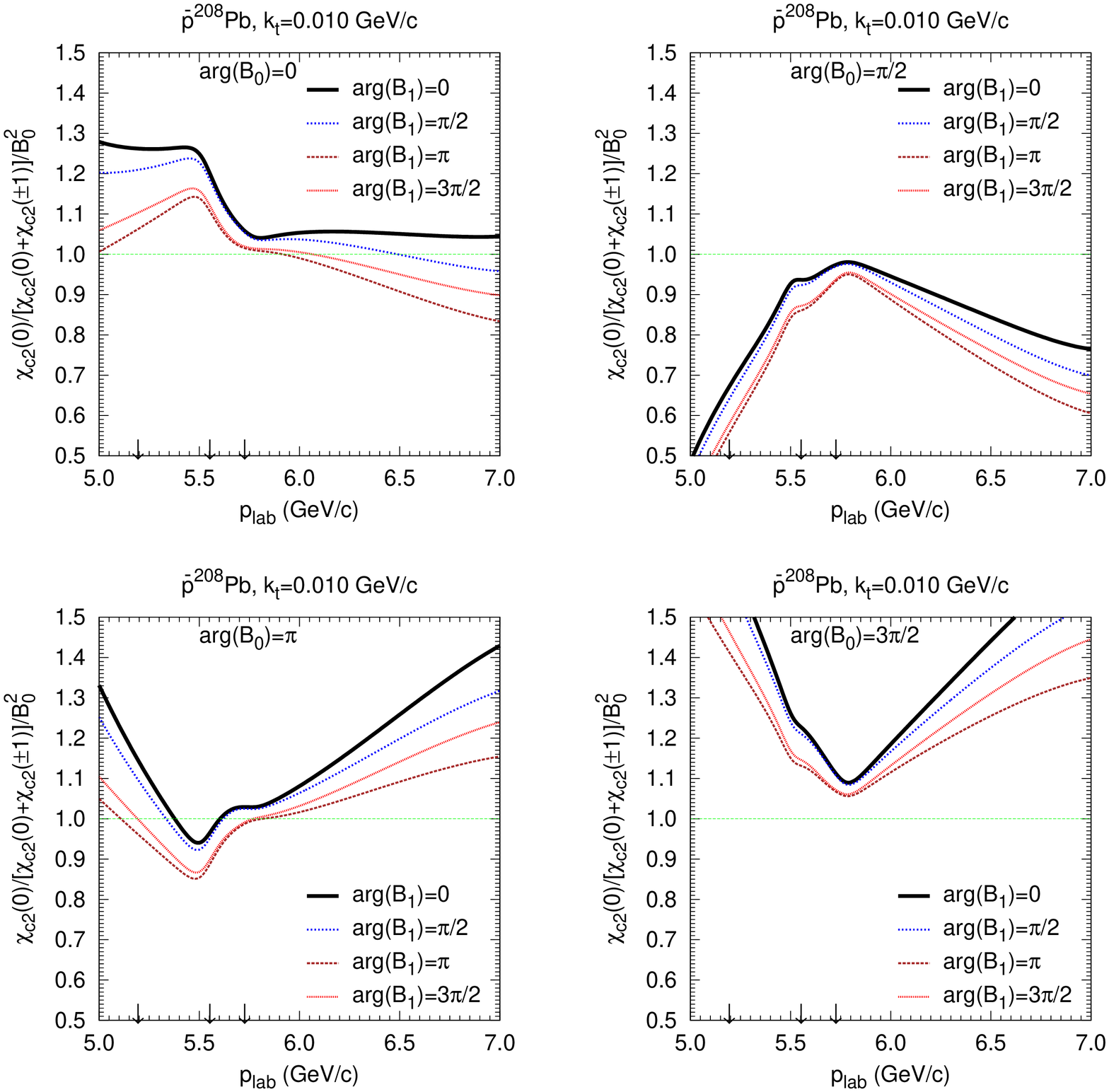}
\caption{\label{fig:ratio_plabDep}(Color online) Same as in 
Fig.~\ref{fig:ratio_plabDep_sum}, but for the different values of
the phases of $B_0$ and $B_1$ amplitudes for $J=2$ as indicated.}
\end{figure}
It is important to note, that all previous results were obtained with 
the zero phases for the $B_0$ and $B_1$ helicity amplitudes for $J=2$. 
Figure~\ref{fig:ratio_plabDep} shows the sensitivity of the ratio ${\cal R}$ 
to the choice of phases for the $B_0$ and $B_1$ amplitudes of $\chi_2$
formation in $\bar p p$ collisions. The $B_0$-phase turns out to be
very important: it governs the shape of the beam momentum
dependence of ${\cal R}$. The $B_1$-phase somewhat shifts ${\cal R}$
vertically but does not influence much the shape of $p_{\rm lab}$-dependence.
This is expected since the direct (leading order) contribution to $\chi_{21}$ 
production is much larger as compared to the direct contribution to $\chi_{20}$
production (compare Figs.~\ref{fig:d2sig_21} and \ref{fig:d2sig_20}).
Hence the interference is relatively less important for $\chi_{21}$. 
 
Finally, we would like to make few comments on the possibility of
experimental measurements of the polarization effects in the $\chi_c$
production at the PANDA@FAIR experiment.    
The PANDA experimental program \cite{PANDA} already includes
the studies of the $\bar p p \to \chi_c \to J/\psi \gamma \to e^+ e^- \gamma$
reaction. The separation of the different $\chi_c$ flavors is possible via the different
energies of the photon \footnote{
In the laboratory frame this obviously corresponds to  a rather broad distribution
over  the photon  energies. Thus, the photon energy should be determined  
in the $e^+ e^- \gamma$ c.m. frame, which gives $E_\gamma=303,~389$ and
430 MeV for $\chi_{c0},~\chi_{c1}$ and $\chi_{c2}$, respectively.
Together with the requirement that the $e^+ e^- \gamma$ 
invariant mass is equal to the respective $\chi_c$ mass this gives
a clean trigger condition of the $\chi_c \to J/\psi \gamma$ decay.}.
This can also be done in the case of nuclear target.

For the $\bar p A$ reactions, the change in the population of the 
low $k_t$ $\chi_{J\nu}$-states with respect to the one for $\bar p p$ reactions
will manifest itself in the change of the polar angle distribution of the 
$J/\psi$-emission for the $\chi_J \to J/\psi \gamma$ decay in the $\chi_J$
rest frame. Neglecting the $\chi_{2,\pm2}$
contribution, this distribution can be expressed as
\begin{equation}
   W_J(\Theta)=\sum_{\nu=\pm1,0} P_{J\nu} W_{J\nu}(\Theta)~,   \label{W_J}
\end{equation}
where
\begin{equation}
   P_{J\nu}=\frac{\chi_{J\nu}}{\chi_{J0}+2\chi_{J1}} \label{P_Jnu}
\end{equation}
is the relative fraction of $\chi_{J\nu}$-states. In particular, for $J=2$,
$P_{20}={\cal R}|B_0|^2$, $P_{2,\pm1}=(1-{\cal R}|B_0|^2)/2$.
The polar angle distribution of $J/\psi$'s in the $\chi_{J\nu}$ radiative decay
is
\begin{equation}
    W_{J\nu}(\Theta) \propto \sum_{\nu^\prime=0}^J |A^J_{\nu^\prime}|^2
                 (  [d^J_{\nu\nu^\prime}(\Theta)]^2 
                  + [d^J_{\nu,-\nu^\prime}(\Theta)]^2 )~.    \label{W_Jnu}
\end{equation}\
This equation includes the helicity amplitudes $A^J_{\nu^\prime}$ 
of the radiative decay which can be further expressed via the amplitudes
$a_1,\ldots,a_{J+1}$ of electric or magnetic multipole transitions such that 
$a_1,~a_2$ and $a_3$ correspond to E1, M2 and E3 transitions \cite{Ridener:1992xr}.
The helicity amplitudes $|B_0|^2$ and $a_2$ for $\chi_{c2}$ are
experimentally known only with an accuracy of about $30-60\%$ from E835 
measurements \cite{Ambrogiani:2001jw}.
Hence it is very important to perform the polarization studies 
{\it within the same experimental setup} not only for $\bar p A$,
but also for $\bar p p$ reaction. Only such parallel measurements could really address 
the nuclear effects discussed in the present work.

\section{Summary}
\label{summary}

We have calculated the transverse momentum differential
cross sections of the polarized $\chi_c$ production in the antiproton-induced reactions
on nuclei close to the production threshold. The incoming antiproton was
assumed to be unpolarized. We have used the multiple scattering
Feynman diagram formalism in the GEA-approach of Refs. 
\cite{Frankfurt:1996xx,Sargsian01}. For the elementary amplitudes
we used expressions motivated by the phenomenology of $\bar p p$-interactions
and QCD. The modifications of the proton occupation numbers due to the short-range
NN correlations in the  nuclear ground state have been taken into account.
The calculated differential cross sections have a characteristic two-slope 
structure. The slope is changed at $k_t \simeq 0.25$ GeV/c due to the SRC-tail
of the proton momentum distribution at high transverse momenta.

As the polarization observable we have chosen the relative fraction ${\cal R}$ of
the $\chi_{c2}$ states with helicity 0 at small transverse momenta normalized
such that ${\cal R}=1$ in the $\bar p p \to \chi_{c2}$ reaction.
The color filtering mechanism alone leads at most to the 10\% increase of
${\cal R}$ in $\bar pA$ reactions with respect to the $\bar p p$ case.  
The interference of the direct $\bar p p \to \chi_{20}$ formation amplitude
with the two-step $\bar p p \to \chi_{00}, \chi_{00} N \to \chi_{20} N$
amplitude strongly influences ${\cal R}$. As a consequence, within a beam 
momentum range of 5-7 GeV/c, ${\cal R}$ varies by 30-50\%. The specific shape
of the $p_{\rm lab}$-dependence of ${\cal R}$ is determined by the unknown 
phase difference of the $B_0$ helicity amplitudes for $J=2$ and $J=0$.
However, the amplitude of the deviations of ${\cal R}$ from the $\bar p p$ 
value is proportional to the difference between the total interaction cross
sections of the 1P charmonium states with $L_z=1$ and $L_z=0$.

To conclude, we suggest that the experimental measurements of the $p_{\rm lab}$
dependence of the relative fraction of $\chi_{c2}$ states with helicity 0
at small transverse momenta in $\bar p A$ reactions would provide a
sensitive test of the constituent quark model description of the $\chi_c$ states  
Such studies can be performed in the future PANDA experiment at FAIR.

\begin{acknowledgments}
M.S. wants to thank Helmholtz Institute in Mainz and GAUSTEQ 
for support during initial stage of work on this project.
We gratefully acknowledge support by the Frankfurt Center for Scientific
Computing. 
This work was supported by HIC for FAIR within the framework of
the LOEWE program (Germany), and by the Grant NSH-215.2012.2 (Russia).  
\end{acknowledgments}

\bibliography{pbarChic}

\appendix

\section{Multiple scattering diagrams}
\label{MultScatt}

\begin{figure}
\includegraphics[scale = 0.7]{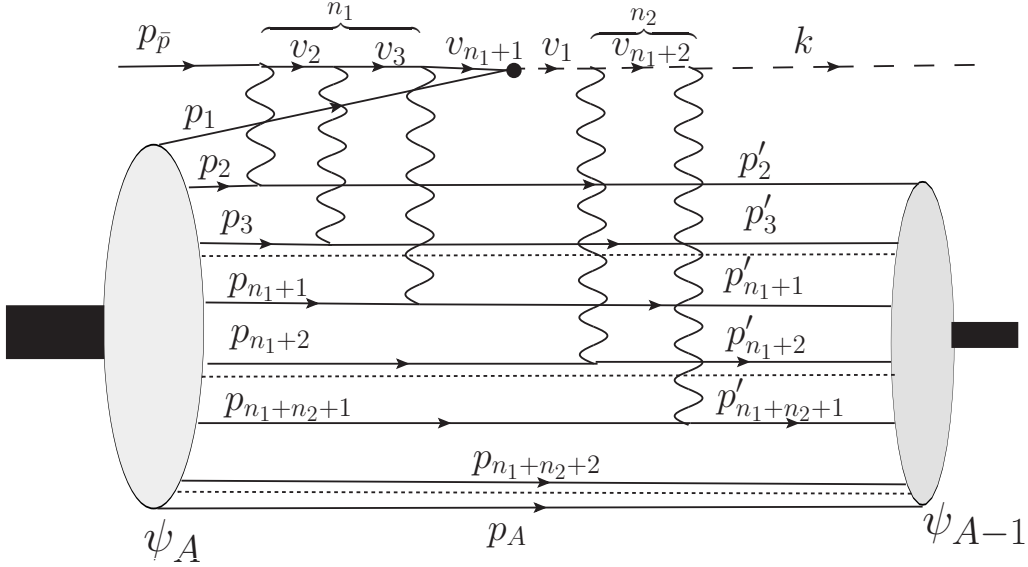}
\caption{\label{fig:pbarChi_diag_mult} The diagonal transition diagram 
for the production of the charmonium state $\chi_J$ (c.f. Fig.~\ref{fig:pbarChi_dia}a)  
including multiple elastic rescatterings for the incoming $\bar p$ 
and outgoing $\chi_J$.}
\end{figure}
The diagonal transition term with with $n_1$ elastic rescatterings
of the antiproton before its annihilation and $n_2$ elastic rescatterings
of the outgoing charmonium $\chi_J$ is shown in Fig.~\ref{fig:pbarChi_diag_mult}.
The full transition amplitude (i.e. $S$-matrix element) between the initial
state antiproton + nucleus $A$  and final state charmonium + nucleus $(A-1)$
is
\begin{eqnarray}
  S_{J\psi_{A-1}; \bar p \psi_{A}} &=&
   \left(\frac{1}{\sqrt{V}}\right)^{2A-1}
   \int d^3x_2^\prime \cdots d^3x_A^\prime \int d^3x_1 \cdots d^3x_A \nonumber \\
   &&\times \psi^*_{A-1}(\mathbf{x}_2^\prime, \ldots ,\mathbf{x}_A^\prime)
            \psi_{A}(\mathbf{x}_1,\mathbf{x}_2, \ldots ,\mathbf{x}_A) \nonumber \\
   &&\times \int \frac{Vd^3p_2^\prime}{(2\pi)^3} \cdots \frac{Vd^3p_n^\prime}{(2\pi)^3}
            \int \frac{Vd^3p_1}{(2\pi)^3} \cdots \frac{Vd^3p_A}{(2\pi)^3} \nonumber \\
   &&\times \mbox{e}^{i\mathbf{p}_2^\prime\mathbf{x}_2^\prime + 
                      \cdots + i\mathbf{p}_n^\prime\mathbf{x}_n^\prime
                      +i\mathbf{p}_{n+1}\mathbf{x}_{n+1}^\prime +
                      \cdots +i\mathbf{p}_{A}\mathbf{x}_{A}^\prime} \nonumber \\
   &&\times  S_{J N_2^\prime \cdots N_n^\prime; \bar p N_1 \cdots N_n}
             \mbox{e}^{-i\mathbf{p}_1\mathbf{x}_1- \cdots -i\mathbf{p}_{A}\mathbf{x}_{A}}~,
                                                      \label{S_matr_A}
\end{eqnarray}
where $n=n_1+n_2+1$ is the number of involved nucleons,
$S_{J N_2^\prime \cdots N_n^\prime; \bar p N_1 \cdots N_n}$ 
is the amplitude of the transition between plane-wave
states, and $V$ is a normalization volume. 
Integrating-out the momenta and coordinates of the spectator nucleons
in the final state gives the following expression
\begin{eqnarray}
  S_{J\psi_{A-1}; \bar p \psi_{A}} &=&
   \left(\frac{1}{\sqrt{V}}\right)^{2n-1}
   \int d^3x_2^\prime \cdots d^3x_n^\prime \int d^3x_1 \cdots d^3x_A \nonumber \\
   &&\times \psi^*_{A-1}(\mathbf{x}_2^\prime,\ldots,\mathbf{x}_n^\prime,
            \mathbf{x}_{n+1},\ldots,\mathbf{x}_A)
            \psi_{A}(\mathbf{x}_1,\ldots,\mathbf{x}_A) \nonumber \\
   &&\times \int \frac{Vd^3p_2^\prime}{(2\pi)^3} \cdots \frac{Vd^3p_n^\prime}{(2\pi)^3}
            \int \frac{Vd^3p_1}{(2\pi)^3} \cdots \frac{Vd^3p_n}{(2\pi)^3} \nonumber \\
   &&\times \mbox{e}^{i\mathbf{p}_2^\prime\mathbf{x}_2^\prime + 
                      \cdots + i\mathbf{p}_n^\prime\mathbf{x}_n^\prime}
             S_{J N_2^\prime \cdots N_n^\prime; \bar p N_1 \cdots N_n}
             \mbox{e}^{-i\mathbf{p}_1\mathbf{x}_1- \cdots -i\mathbf{p}_{n}\mathbf{x}_{n}}~.
                                                      \label{S_matr_A_simpl}
\end{eqnarray}
The $S$-matrix element between plane-wave states is expressed as follows
\begin{eqnarray}
    S_{J N_2^\prime \cdots N_n^\prime; \bar p N_1 \cdots N_n} &=&
    i (2\pi)^4 \delta^{(4)}( p_{\bar p}+p_1+p_2+ \cdots +p_n
                            -k-p_2^\prime- \cdots -p_n^\prime ) \nonumber \\
   &&\times \frac{M_{J N_2^\prime \cdots N_n^\prime; \bar p N_1 \cdots N_n}}%
                 {[2E_{\bar p}V 2E_1V (2mV)^{2(n-1)} 2 \omega V]^{1/2}}~,
                \label{S_matr_expl}
\end{eqnarray}
where we assumed that the initial and final nucleons are nonrelativistic.
For the following it is convenient to introduce the four-momentum transfer
by the $i$-th nucleon as
\begin{equation}
   q_i=p_i-p_i^\prime~,~~i=2,\ldots,n~.  \label{4momentum_transfer}
\end{equation}
The corresponding transverse and longitudinal momentum transfers
are then $\mathbf{t}_i \equiv \mathbf{q}_{it}=\mathbf{p}_{it}-\mathbf{p}_{it}^\prime$
and $q_i^z=p_i^z-p_i^{\prime z}$.
 
The invariant amplitude is calculated with a help of Feynman rules
which gives
\begin{eqnarray}
  M_{J N_2^\prime \cdots N_n^\prime; \bar p N_1 \cdots N_n}
  &=& \frac{M_{JN}(\mathbf{t}_n) M_{JN}(\mathbf{t}_{n-1})  
    \cdots M_{JN}(\mathbf{t}_{n_1+2}) M_{J;\bar p p}(\mathbf{p}_{1t})}%
    {D_J(v_{n-1}) \cdots D_J(v_{n_1+2}) D_J(v_{1})} \nonumber \\
  &&\times \frac{M_{\bar p N}(\mathbf{t}_{n_1+1}) 
                 \cdots M_{\bar p N}(\mathbf{t}_2)}%
                {D_{\bar p}(v_{n_1+1}) \cdots D_{\bar p}(v_{2})}~.
                                                         \label{M_elem}
\end{eqnarray}
We assumed here that the elementary amplitudes depend on the transverse momentum
transfers only.
The antiproton inverse propagators are
\begin{equation}
   -D_{\bar p}(v_{i})=(p_{\bar p}+\sum_{j=2}^i q_j)^2
    -m^2 + i\varepsilon = 2p_{\rm lab}(-l_i+i\varepsilon)~,~~i=2,\ldots, n_1+1~.   \label{pbar_prop}
\end{equation}
The charmonium inverse propagators are
\begin{eqnarray}
   -D_{J}(v_{1})&=& (p_{\bar p}+p_1+\sum_{j=2}^{n_1+1} q_j)^2
    -m_J^2 + i\varepsilon = 2p_{\rm lab}(\Delta^0_J-l_1+i\varepsilon)~, \nonumber \\
   -D_{J}(v_{i})&=& (p_{\bar p}+p_1+\sum_{j=2}^{i} q_j)^2
    -m_J^2 + i\varepsilon = 2p_{\rm lab}(\Delta^0_J-l_i+i\varepsilon)~, \label{charm_prop}
\end{eqnarray}
where $i=n_1+2,\ldots,n-1$.
In Eqs.(\ref{pbar_prop}),(\ref{charm_prop}) we used the accumulated
longitudinal momentum transfers defined as
\begin{equation}
   l_i=\left\{ \begin{array}{lll}
               \sum_{j=2}^i q_j^z & \mbox{for} & i=2,\ldots, n_1+1~, \\
              p_1^z+\sum_{j=2}^{n_1+1} q_j^z & \mbox{for} & i=1~, \\
              p_1^z+\sum_{j=2}^{i} q_j^z & \mbox{for} & i=n_1+2,\ldots,n-1~. 
               \end{array}
       \right.
     \label{l_i}
\end{equation}

By using the coordinate representation of the propagators (\ref{Dcoord})
we can now perform the longitudinal momentum integrations in (\ref{S_matr_A_simpl}).
After some algebra we come to the following expression:
\begin{eqnarray}
     \int &dp_2^{\prime z}& \cdots dp_n^{\prime z} 
     \int dp_1^z \cdots dp_n^z
     \delta(p_{\rm lab}+p_1^z+q_2^z+\cdots+q_n^z-k^z) \nonumber \\
  && \times \exp\{ ip_2^{\prime z} z_2^\prime+ \cdots + ip_n^{\prime z} z_n^\prime
           -il_2z_2^0-\cdots-il_{n_1+1}z_{n_1+1}^0  \nonumber \\
  &&\hspace{1.5cm}  + i(\Delta^0_J-l_1)z_1^0+i(\Delta^0_J-l_{n_1+2})z_{n_1+2}^0+
             \cdots +i(\Delta^0_J-l_{n_1+n_2})z_{n_1+n_2}^0  \nonumber \\
  &&\hspace{1.5cm} -i p_1^z z_1- \cdots - i p_n^z z_n\} \nonumber \\
  &=& (2\pi)^{n-1} \delta(z_2^\prime-z_2) \cdots \delta(z_n^\prime-z_n)
                   \exp\{i \Delta^0_J (z_1^0+z_{n_1+2}^0+ \cdots +z_{n_1+n_2}^0)\} \nonumber \\
  && \times \int dq_2^z \cdots dq_n^z 
            \exp\{-i q_2^z z_2- \cdots -i q_n^z z_n
                  -i l_1 z_1^0 - i l_2 z_2^0 - \cdots -i l_{n_1+n_2} z_{n_1+n_2}^0 \nonumber \\
  &&\hspace{3.5cm} +i(p_{\rm lab}-k^z+q_2^z+ \cdots +q_n^z)z_1\}  \nonumber \\
  &=& (2\pi)^{2(n-1)} \delta(z_2^\prime-z_2) \cdots \delta(z_n^\prime-z_n) 
            \delta(z_2-z_3+z_2^0) \cdots \delta(z_{n_1}-z_{n_1+1}+z_{n_1}^0) \nonumber \\
  && \times  \delta(z_{n_1+1}-z_1+z_{n_1+1}^0) \delta(z_{n_1+2}-z_1-z_1^0) \nonumber \\
  && \times  \delta(z_{n_1+3}-z_{n_1+2}-z_{n_1+2}^0) \cdots
             \delta(z_{n}-z_{n-1}-z_{n-1}^0)    \nonumber \\
  && \times  \exp\{i (p_{\rm lab}-k^z+\Delta_J^0) z_n - i \Delta_J^0 z_1\}~.  \label{long_mom_integr}
\end{eqnarray}
In order to obtain the last expression in (\ref{long_mom_integr}) we substituted 
the expression $p_1^z=k^z-p_{\rm lab}-q_2^z- \cdots - q_n^z$ in the formulas (\ref{l_i})
for the accumulated longitudinal momentum transfers and, after performing the integrations
over $dq_2^z \cdots dq_n^z$, simplified the arguments of $\delta$-functions by using
recursive relations 
\begin{eqnarray}
  && z_i+z_i^0+ \cdots + z_{n_1+1}^0 - z_1 = z_i - z_{i+1} + z_i^0~,~~i=n_1,\ldots,2~;  \nonumber \\
  && z_i-z_1^0-z_{n_1+2}^0- \cdots -z_{i-1}^0-z_1=z_i-z_{i-1}-z_{i-1}^0~,~~i=n_1+3,\ldots,n~. \label{rec_rel}
\end{eqnarray}

Transverse momentum integrations in (\ref{S_matr_A_simpl}) are performed as follows:
\begin{eqnarray}
  \int &d^2p_{2t}^\prime& \cdots d^2p_{nt}^\prime 
   \int d^2p_{1t} \cdots d^2p_{nt} 
   \delta^{(2)}(\mathbf{p}_{1t}+\mathbf{t}_2+ \cdots +\mathbf{t}_n - \mathbf{k}_t) \nonumber \\
  && \times M_{\bar p N}(\mathbf{t}_2) \cdots M_{\bar p N}(\mathbf{t}_{n_1+1})
     M_{J;\bar p N}(\mathbf{p}_{1t})
     M_{JN}(\mathbf{t}_{n_1+2}) \cdots M_{JN}(\mathbf{t}_n) \nonumber \\
  && \times \exp\{ i\mathbf{p}_{2t}^\prime \mathbf{b}_{2}^\prime + \cdots
                  +i\mathbf{p}_{nt}^\prime \mathbf{b}_{n}^\prime
                  -i\mathbf{p}_{1t} \mathbf{b}_{1}- \cdots 
                  -i\mathbf{p}_{nt} \mathbf{b}_{n}\}  \nonumber \\
  &=& (2\pi)^{2(n-1)} \delta^{(2)}(\mathbf{b}_{2}^\prime-\mathbf{b}_{2}) \cdots
                      \delta^{(2)}(\mathbf{b}_{n}^\prime-\mathbf{b}_{n})
      \exp(-i \mathbf{k}_t \mathbf{b}_{1})
      \int d^2t_2 \cdots d^2t_n   \nonumber \\ 
 && \times \exp\{ -i \mathbf{t}_2 (\mathbf{b}_{2}-\mathbf{b}_{1})
                  - \cdots 
                  -i \mathbf{t}_n (\mathbf{b}_{n}-\mathbf{b}_{1})
                \}
     M_{\bar p N}(\mathbf{t}_2) \cdots M_{\bar p N}(\mathbf{t}_{n_1+1}) \nonumber \\
 && \times M_{J;\bar p p}(\mathbf{k}_t-\mathbf{t}_2- \cdots -\mathbf{t}_n)
     M_{JN}(\mathbf{t}_{n_1+2}) \cdots M_{JN}(\mathbf{t}_n)~.  \label{trans_mom_integr}
\end{eqnarray}
We see that due to the $\delta$-functions in Eqs.(\ref{long_mom_integr}),(\ref{trans_mom_integr})
the primed and nonprimed coordinates coincide and the integration over  
$d^3x_2^\prime \cdots d^3x_n^\prime$ in  Eq. (\ref{S_matr_A_simpl}) leads
to the appearance of the product $\psi^*_{A-1}(\mathbf{x}_2,\ldots,\mathbf{x}_A)
\psi_{A}(\mathbf{x}_1,\ldots,\mathbf{x}_A)$ in the transition amplitude. 

Using Eqs.(\ref{long_mom_integr}),(\ref{trans_mom_integr}) 
and assuming again
the nonrelativistic nucleons (i.e. neglecting the energy transfer in rescattering processes)
we can rewrite the amplitude (\ref{S_matr_A_simpl}) as
\begin{equation}
   S_{J\psi_{A-1}; \bar p \psi_{A}} = 
   \frac{i(2\pi) \delta(E_{\bar p}+E_1-\omega)}{(2E_{\bar p}V 2\omega V)^{1/2}}
   M_{J\psi_{A-1}; \bar p \psi_{A}}~,    \label{S_matr_def}
\end{equation}
where the matrix element $M_{J\psi_{A-1}; \bar p \psi_{A}}$ should be replaced 
by the following one:
\begin{eqnarray}
  M^J(1,2,\ldots,n) &=& 
  \frac{i^{n-1}}{(2E_1)^{1/2} (2\pi)^{2(n-1)} (4mp_{\rm lab})^{n-1}}  \nonumber \\
  && \times  \int d^3x_1 \cdots d^3x_A
            \psi^*_{A-1}(\mathbf{x}_2,\ldots,\mathbf{x}_A)
            \psi_{A}(\mathbf{x}_1,\ldots,\mathbf{x}_A) \nonumber \\
  && \times \Theta(z_3-z_2) \cdots \Theta(z_{n_1+1}-z_{n_1}) 
            \Theta(z_1-z_{n_1+1}) \Theta(z_{n_1+2}-z_1)       \nonumber \\
  && \times  \Theta(z_{n_1+3}-z_{n_1+2}) \cdots
             \Theta(z_{n}-z_{n-1})    \nonumber \\
  && \times \exp\{i (p_{\rm lab}-k^z+\Delta_J^0) z_n - i \Delta_J^0 z_1 
       -i \mathbf{k}_t \mathbf{b}_{1}\}  
      \int d^2t_2 \cdots d^2t_n   \nonumber \\ 
 && \times \exp\{ -i \mathbf{t}_2 (\mathbf{b}_{2}-\mathbf{b}_{1})
                  - \cdots 
                  -i \mathbf{t}_n (\mathbf{b}_{n}-\mathbf{b}_{1})
                \}
     M_{\bar p N}(\mathbf{t}_2) \cdots M_{\bar p N}(\mathbf{t}_{n_1+1}) \nonumber \\
 && \times M_{J;\bar p p}(\mathbf{k}_t-\mathbf{t}_2- \cdots -\mathbf{t}_n)
     M_{JN}(\mathbf{t}_{n_1+2}) \cdots M_{JN}(\mathbf{t}_n)~.    \label{M^Jmult_fixedOrder}
\end{eqnarray}
The product of $\Theta$-functions in this equation is governed by 
the order of scatterings of the incoming antiproton and outgoing charmonium.
Hence, summing all possible diagrams with the different order of scatterings
on the fixed sets of nucleons-scatterers ($n_1$ scatterers for the $\bar p$ and 
$n_2$ scatterers the for charmonium) is equivalent to the replacement
of the product of the $\Theta$-functions in (\ref{M^Jmult_fixedOrder}) by
the following one:
\begin{equation}
      \Theta(z_1-z_2) \cdots \Theta(z_1-z_{n_1+1})  
      \Theta(z_{n_1+2}-z_1) \cdots \Theta(z_{n}-z_1)~. \label{Theta_product_diag}
\end{equation}
Let us now constrain the kinematics of the produced charmonium such that
$|p_{\rm lab}-k^z| \ll p_{\rm lab}$, i.e to the quasifree region.    
Due to the presence of 
$\delta(E_{\bar p}+E_1-\omega)$ in the expressions for the $S$-matrix (\ref{S_matr_def})
and in the differential cross section (\ref{dsigma}), such a constraint leads
to the condition
\begin{equation}
    p_{\rm lab}-k^z+\Delta_J^0 \simeq (\Delta_J^0)^2/2p_{\rm lab} \ll \Delta_J^0~. \label{kin_constr}
\end{equation}
And thus we can neglect the term $i (p_{\rm lab}-k^z+\Delta_J^0) z_n$ 
in the exponent of Eq.(\ref{M^Jmult_fixedOrder})
which depends on the longitudinal coordinate $z_n$ of the last scatterer.
This leads us to the following expression for the matrix element of the
diagonal transition with multiple elastic rescatterings:
\begin{eqnarray}
  M^J(1,2,\ldots,n) &=& 
  \frac{i^{n-1}}{(2E_1)^{1/2} (2\pi)^{2(n-1)} (4mp_{\rm lab})^{n-1}}
      \int d^3x_1 \cdots d^3x_A   \nonumber \\
   && \times \psi^*_{A-1}(\mathbf{x}_2,\ldots,\mathbf{x}_A)
             \psi_{A}(\mathbf{x}_1,\ldots,\mathbf{x}_A)
               \Theta(z_1-z_2) \cdots \Theta(z_1-z_{n_1+1}) \nonumber \\
  && \times    \Theta(z_{n_1+2}-z_1) \cdots \Theta(z_{n}-z_1)
     \exp\{-i \Delta_J^0 z_1 -i \mathbf{k}_t \mathbf{b}_{1}\}
     \int d^2t_2 \cdots d^2t_n  \nonumber \\  
  && \times \exp\{ -i \mathbf{t}_2 (\mathbf{b}_{2}-\mathbf{b}_{1})
                  - \cdots 
                  -i \mathbf{t}_n (\mathbf{b}_{n}-\mathbf{b}_{1})
                \}       
      M_{\bar p N}(\mathbf{t}_2) \cdots M_{\bar p N}(\mathbf{t}_{n_1+1}) \nonumber \\
  && \times M_{J;\bar p p}(\mathbf{k}_t-\mathbf{t}_2- \cdots -\mathbf{t}_n)
     M_{JN}(\mathbf{t}_{n_1+2}) \cdots M_{JN}(\mathbf{t}_n)~.    \label{M^Jmult}
\end{eqnarray}  

\begin{figure}
\includegraphics[scale = 0.5]{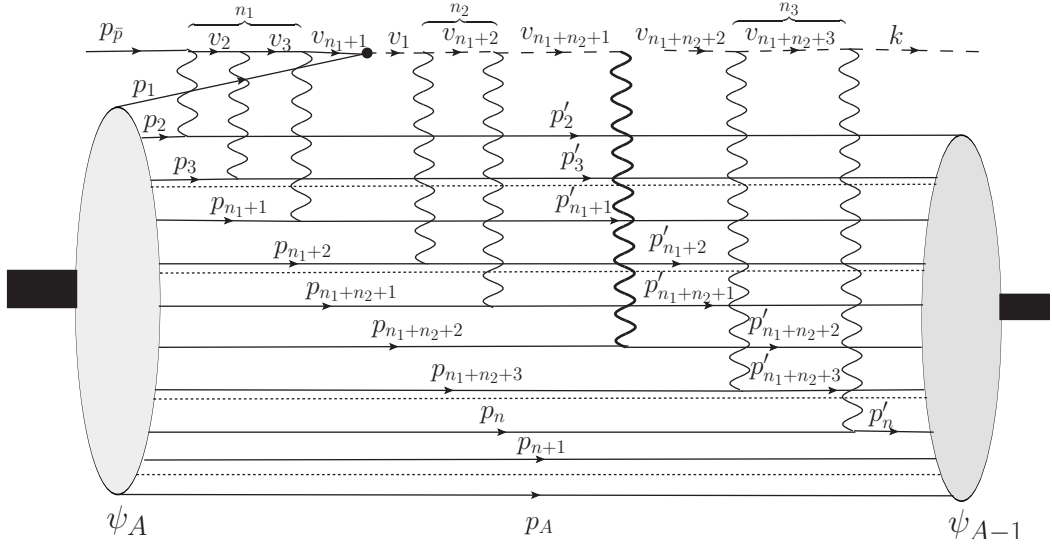}
\caption{\label{fig:pbarChi_nondiag_mult} The diagram with one 
nondiagonal transition $\chi_{J_1} N_{n_1+n_2+2} 
\to \chi_{J} N_{n_1+n_2+2}^\prime$ (c.f. Fig.~\ref{fig:pbarChi_dia}c)  
including multiple elastic rescatterings for the incoming $\bar p$, intermediate
$\chi_{J_1}$ and outgoing $\chi_J$.}
\end{figure}
The diagram with one nondiagonal transition, $n_1$ elastic rescatterings 
of the incoming antiproton, $n_2$ elastic rescatterings of intermediate
charmonium $\chi_{J_1}$ and $n_3$ elastic rescatterings of the outgoing
charmonium $\chi_{J}$ is shown in Fig.~\ref{fig:pbarChi_nondiag_mult}.
In total, $n=n_1+n_2+n_3+2$ nucleons are involved in the reaction.
It is clear then, that the formulas ({\ref{S_matr_A})-(\ref{4momentum_transfer})
are valid also in this case, but with the newly defined value of $n$.
For the invariant amplitude we have now instead of Eq.(\ref{M_elem}):
\begin{eqnarray}
  M_{J N_2^\prime \cdots N_n^\prime; \bar p N_1 \cdots N_n}
  &=& \frac{M_{JN}(\mathbf{t}_n) M_{JN}(\mathbf{t}_{n-1})  
            \cdots M_{JN}(\mathbf{t}_{n_1+n_2+3})
                   M_{JN;J_1N}(\mathbf{t}_{n_1+n_2+2})}%
           {D_J(v_{n-1}) \cdots D_J(v_{n_1+n_2+2})} \nonumber \\
  && \times \frac{M_{J_1N}(\mathbf{t}_{n_1+n_2+1})
            \cdots M_{J_1N}(\mathbf{t}_{n_1+2})
                 M_{J_1;\bar p p}(\mathbf{p}_{1t})}%
           {D_{J_1}(v_{n_1+n_2+1}) \cdots D_{J_1}(v_{n_1+2}) D_{J_1}(v_{1})} \nonumber \\
  && \times \frac{M_{\bar p N}(\mathbf{t}_{n_1+1})
                 \cdots M_{\bar p N}(\mathbf{t}_2)}%            
           {D_{\bar p}(v_{n_1+1}) \cdots D_{\bar p}(v_{2})}~.
                                                         \label{M_elem_nondiag}
\end{eqnarray}
The antiproton inverse propagators are given by Eq.(\ref{pbar_prop}).
The $\chi_{J_1}$-charmonium inverse propagators are
\begin{eqnarray}
   -D_{J_1}(v_{1})&=& (p_{\bar p}+p_1+\sum_{j=2}^{n_1+1} q_j)^2
    -m_{J_1}^2 + i\varepsilon = 2p_{\rm lab}(\Delta^0_{J_1}-l_1+i\varepsilon)~, \nonumber \\
   -D_{J_1}(v_{i})&=& (p_{\bar p}+p_1+\sum_{j=2}^{i} q_j)^2
    -m_{J_1}^2 + i\varepsilon = 2p_{\rm lab}(\Delta^0_{J_1}-l_i+i\varepsilon)~, \label{charm1_prop_nond}
\end{eqnarray}
where $i=n_1+2,\ldots,n_1+n_2+1$. 
The $\chi_{J}$-charmonium inverse propagators are
\begin{eqnarray}
    -D_{J}(v_{i})&=& (p_{\bar p}+p_1+\sum_{j=2}^{i} q_j)^2
    -m_{J}^2 + i\varepsilon = 2p_{\rm lab}(\Delta^0_{J}-l_i+i\varepsilon)~, \label{charm_prop_nond}
\end{eqnarray}    
where $i=n_1+n_2+2,\ldots,n-1$. The accumulated longitudinal momentum transfers $l_i$ are given by 
Eqs.(\ref{l_i}) with the new value of $n$.

The longitudinal momentum integration in (\ref{S_matr_A_simpl}) becomes now
\begin{eqnarray}
  \int &dp_2^{\prime z}& \cdots dp_n^{\prime z} 
     \int dp_1^z \cdots dp_n^z
     \delta(p_{\rm lab}+p_1^z+q_2^z+\cdots+q_n^z-k^z) \nonumber \\
  && \times \exp\{ ip_2^{\prime z} z_2^\prime+ \cdots + ip_n^{\prime z} z_n^\prime
           -il_2z_2^0-\cdots-il_{n_1+1}z_{n_1+1}^0  \nonumber \\
  &&\hspace{1.5cm}  + i(\Delta^0_{J_1}-l_1)z_1^0+i(\Delta^0_{J_1}-l_{n_1+2})z_{n_1+2}^0+
             \cdots +i(\Delta^0_{J_1}-l_{n_1+n_2+1})z_{n_1+n_2+1}^0  \nonumber \\
  &&\hspace{1.5cm}  + i(\Delta^0_{J}-l_{n_1+n_2+2})z_{n_1+n_2+2}^0 
      + \cdots +i(\Delta^0_{J}-l_{n-1})z_{n-1}^0
      -i p_1^z z_1- \cdots - i p_n^z z_n\}        \nonumber \\
  &=& (2\pi)^{n-1} \delta(z_2^\prime-z_2) \cdots \delta(z_n^\prime-z_n)   \nonumber \\
  && \times \exp\{  i \Delta^0_{J_1} (z_1^0+z_{n_1+2}^0+ \cdots +z_{n_1+n_2+1}^0)
                  + i \Delta^0_{J} (z_{n_1+n_2+2}^0 + \cdots +z_{n-1}^0) \}  \nonumber \\
  && \times \int dq_2^z \cdots dq_n^z 
            \exp\{-i q_2^z z_2- \cdots -i q_n^z z_n
                  -i l_1 z_1^0 - \cdots -i l_{n-1} z_{n-1}^0 \nonumber \\
  &&\hspace{3.5cm}             +i(p_{\rm lab}-k^z+q_2^z+ \cdots +q_n^z)z_1\}  \nonumber \\
  &=& (2\pi)^{2(n-1)} \delta(z_2^\prime-z_2) \cdots \delta(z_n^\prime-z_n) 
            \delta(z_2-z_3+z_2^0) \cdots \delta(z_{n_1}-z_{n_1+1}+z_{n_1}^0) \nonumber \\
  && \times  \delta(z_{n_1+1}-z_1+z_{n_1+1}^0) \delta(z_{n_1+2}-z_1-z_1^0) \nonumber \\
  && \times  \delta(z_{n_1+3}-z_{n_1+2}-z_{n_1+2}^0) \cdots
             \delta(z_{n}-z_{n-1}-z_{n-1}^0)    \nonumber \\
  && \times  \exp\{ i (p_{\rm lab}-k^z+\Delta_J^0) z_n 
                  - i \Delta_{J_1}^0 z_1 
                  + i (\Delta_{J_1}^0-\Delta_{J}^0)z_{n_1+n_2+2} \}~.  \label{long_mom_integr_nond}
\end{eqnarray}
The derivation of the last expression in (\ref{long_mom_integr_nond}) was performed 
in full analogy with the case of the longitudinal integral (\ref{long_mom_integr})
for the diagonal amplitude. We again used the formulas (\ref{l_i}) for the accumulated longitudinal
momentum transfers with $p_1^z=k^z-p_{\rm lab}-q_2^z- \cdots -q_n^z$ and applied
the recursive relations (\ref{rec_rel}) in the arguments of the $\delta$-functions
(with newly defined $n=n_1+n_2+n_3+2$).

Transverse momentum integral in (\ref{S_matr_A_simpl}) for the nondiagonal amplitude 
has the following form:
\begin{eqnarray}
  \int &d^2p_{2t}^\prime& \cdots d^2p_{nt}^\prime 
   \int d^2p_{1t} \cdots d^2p_{nt} 
   \delta^{(2)}(\mathbf{p}_{1t}+\mathbf{t}_2+ \cdots +\mathbf{t}_n - \mathbf{k}_t)
    M_{\bar p N}(\mathbf{t}_2) \cdots M_{\bar p N}(\mathbf{t}_{n_1+1}) \nonumber \\ 
  && \times M_{J_1;\bar p p}(\mathbf{p}_{1t})
     M_{J_1N}(\mathbf{t}_{n_1+2}) \cdots M_{J_1N}(\mathbf{t}_{n_1+n_2+1}) 
     M_{JN;J_1N}(\mathbf{t}_{n_1+n_2+2})                \nonumber \\            
  && \times M_{JN}(\mathbf{t}_{n_1+n_2+3}) \cdots M_{JN}(\mathbf{t}_{n})
            \exp\{ i\mathbf{p}_{2t}^\prime \mathbf{b}_{2}^\prime + \cdots
                  +i\mathbf{p}_{nt}^\prime \mathbf{b}_{n}^\prime
                  -i\mathbf{p}_{1t} \mathbf{b}_{1}- \cdots 
                  -i\mathbf{p}_{nt} \mathbf{b}_{n}\}  \nonumber \\
  &=& (2\pi)^{2(n-1)} \delta^{(2)}(\mathbf{b}_{2}^\prime-\mathbf{b}_{2}) \cdots
                      \delta^{(2)}(\mathbf{b}_{n}^\prime-\mathbf{b}_{n})
      \exp(-i \mathbf{k}_t \mathbf{b}_{1})
      \int d^2t_2 \cdots d^2t_n   \nonumber \\ 
 && \times \exp\{ -i \mathbf{t}_2 (\mathbf{b}_{2}-\mathbf{b}_{1})
                  - \cdots 
                  -i \mathbf{t}_n (\mathbf{b}_{n}-\mathbf{b}_{1})
                \}
     M_{\bar p N}(\mathbf{t}_2) \cdots M_{\bar p N}(\mathbf{t}_{n_1+1}) \nonumber \\
 && \times  M_{J_1;\bar p p}(\mathbf{k}_t-\mathbf{t}_2- \cdots -\mathbf{t}_n) 
            M_{J_1N}(\mathbf{t}_{n_1+2}) \cdots M_{J_1N}(\mathbf{t}_{n_1+n_2+1}) \nonumber \\ 
 && \times  M_{JN^\prime;J_1N}(\mathbf{t}_{n_1+n_2+2}) 
            M_{JN}(\mathbf{t}_{n_1+n_2+3}) \cdots M_{JN}(\mathbf{t}_{n})~.  \label{trans_mom_integr_nond}
\end{eqnarray}
Using (\ref{long_mom_integr_nond}),(\ref{trans_mom_integr_nond}) we can 
express the amplitude of the nondiagonal transition including multiple elastic 
rescatterings in the form (\ref{S_matr_def}) with the matrix element
$M_{J\psi_{A-1}; \bar p \psi_{A}}$ replaced by  
\begin{eqnarray}
  M^{J_1J}(1,2,\ldots,n) &=& 
      \frac{i^{n-1}}{(2E_1)^{1/2} (2\pi)^{2(n-1)} (4mp_{\rm lab})^{n-1}}
              \int d^3x_1 \cdots d^3x_A  \nonumber \\
  && \times   \psi^*_{A-1}(\mathbf{x}_2,\ldots,\mathbf{x}_A)
              \psi_{A}(\mathbf{x}_1,\ldots,\mathbf{x}_A)
              \Theta(z_1-z_2) \cdots \Theta(z_1-z_{n_1+1}) \nonumber \\
  && \times \Theta(z_{n_1+2}-z_1)\Theta(z_{n_1+n_2+2}-z_{n_1+2})
            \cdots
            \Theta(z_{n_1+n_2+1}-z_1)\Theta(z_{n_1+n_2+2}-z_{n_1+n_2+1}) \nonumber \\
  && \times \Theta(z_{n_1+n_2+3}-z_{n_1+n_2+2}) \cdots \Theta(z_{n}-z_{n_1+n_2+2}) \nonumber \\ 
  && \times \exp\{-i \Delta_{J_1}^0 z_1 -i \mathbf{k}_t \mathbf{b}_{1}
                  +i (\Delta_{J_1}^0-\Delta_{J}^0)z_{n_1+n_2+2}\} 
            \int d^2t_2 \cdots d^2t_n    \nonumber \\    
  && \times \exp\{ -i \mathbf{t}_2 (\mathbf{b}_{2}-\mathbf{b}_{1})
                  - \cdots 
                  -i \mathbf{t}_n (\mathbf{b}_{n}-\mathbf{b}_{1})
                \}                    
      M_{\bar p N}(\mathbf{t}_2) \cdots M_{\bar p N}(\mathbf{t}_{n_1+1}) \nonumber \\
  && \times M_{J_1;\bar p p}(\mathbf{k}_t-\mathbf{t}_2- \cdots -\mathbf{t}_n)
      M_{J_1N}(\mathbf{t}_{n_1+2}) \cdots M_{J_1N}(\mathbf{t}_{n_1+n_2+1}) \nonumber \\    
  && \times  M_{JN^\prime;J_1N}(\mathbf{t}_{n_1+n_2+2})
      M_{JN}(\mathbf{t}_{n_1+n_2+3}) \cdots M_{JN}(\mathbf{t}_n)~,  \label{M^J1Jmult}
\end{eqnarray}
where we summed the diagrams with the different order of rescatterings
(with the fixed struck nucleon $N_1$ and nucleon $N_{n_1+n_2+2}$ on which the
nondiagonal transition takes place) and made the assumption of the
quasifree kinematics $|p_{\rm lab}-k^z| \ll p_{\rm lab}$  of the final $\chi_J$.
Equations (\ref{M^Jmult}),(\ref{M^J1Jmult}) are the generalizations 
of the corresponding Eqs.(\ref{M^J(1)}),(\ref{M^J1J(1,2)}) for the case of 
multiple elastic rescatterings of the antiproton and charmonia.

\end{document}